\documentclass[aps,groupedaddress,notitlepage,longbibliography,nofootinbib,showkeys,showpacs,amssymb]{revtex4-1}
\usepackage{latexsym}
\usepackage{amsfonts}
\usepackage{mathptmx}      % use Times fonts if available on your TeX system
\usepackage{bm}% bold math symbols \bm\alpha
\usepackage{enumerate}
\usepackage{color}
\usepackage{ulem}
\usepackage[dvipdfmx]{graphicx}

\usepackage{amsmath,amssymb}
\usepackage{float}
{%
   \begin{list}{$\bullet$\ \ }%
   {%
      \setlength{\itemindent}{0pt}
      \setlength{\leftmargin}{0.5zw}%
      \setlength{\rightmargin}{0zw}%
      \setlength{\labelsep}{0.5zw}%
      \setlength{\labelwidth}{0zw}%
      \setlength{\itemsep}{0em}%
      \setlength{\parsep}{0em}%
      \setlength{\listparindent}{0zw}%
   }
}{%
   \end{list}%
}

\newcommand{\ignore}[1]{}
\newcommand{\beq}{\begin{equation}}
\newcommand{\eeq}{\end{equation}}

\begin{document}
\title
{Temperature- and Size-dependence of Line shape of ESR spectra of XXZ antiferromagnetic chain}

\author{Hiroki Ikeuchi$^{1}$}
%\email[Corresponding author. Email address: ]{hiroki.ikeuch@gmail.com}
\author{Hans De Raedt$^{2}$}
%\email[Corresponding author. Email address: ]{ikeuch@spin.phys.s.u-tokyo.ac.jp}
\author{Sylvain Bertaina$^{3}$}
\author{Seiji Miyashita$^{1}$}
\email[Corresponding author. Email address: ]{miyashita@phys.s.u-tokyo.ac.jp}
%\email[Corresponding author. Email address: ]{miyashita@phys.s.u-tokyo.ac.jp}

\affiliation{$^{1}
${\it Department of Physics, Graduate School of Science,} The University of Tokyo, 7-3-1 Bunkyo-Ku, Tokyo, 113-0033, Japan \\
$^{2}${\it Department of Applied Physics, Zernike Institute for Advanced Materials,
University of Groningen, Nijenborgh 4, NL-9747AG Groningen, The Netherlands}\\
$^{3}${\it Aix-Marseille Universit\'e, CNRS, IM2NP UMR7334, F-13397 Marseille Cedex 20, France}
}

\date{\today}

\begin{abstract}
The ESR (Electron Spin Resonance) spectrum of the XXZ spin chain with finite length shows a double-peak structure at high temperatures around the EPR (Electron Paramegnetic Resonance) resonance frequency. This fact has been pointed out by direct numerical methods 
(S. El Shawish, O. C\'epas and S. Miyashita: Phys. Rev. B \textbf{81}, 224421 (2010); 
H. Ikeuchi, H. De Raedt, S. Bertaina, and S. Miyashita: Phys. Rev. B \textbf{92}, 214431 (2015)). 
On the other hand, at low temperatures. the spectrum has a single peak with a finite shift from 
%miya0724
the frequency of EPR
as predicted by the analysis of field theoretical works (M. Oshikawa and I. Affleck:
Phys. Rev. Lett. \textbf{82}, 5136 (1999)).
We study how the spectrum changes with the temperature, and also we study the size-dependence of the line shape including the even-odd effect. In order to understand those dependences, we introduce a decomposition of the spectrum into contributions from transitions specified by magnetization, and we characterize the structure of the spectrum by individual contributions. Applying the moment method introduced by M. Brockman et al., to each component, we analyze the size-dependence of the structure of the spectrum, which supports the numerical observation that separation of the double-peak structure vanishes inversely with the size.
\end{abstract}

\pacs{05.30.-d,75.10.jm,76.30.-v}

\keywords{ESR, XXZ model, line shape}%Use showkeys class option if keyword
                              %display desired
\maketitle

%----------------------------------------------------------------------------
\section{Introduction}

The ESR (Electron Spin Resonance) is one of the major tools to obtain information about the spin ordering.
To understand the spectrum, parameter dependence of a concrete ESR spectrum for a specified system has to be clarified, including the temperature-dependence.
To study these aspects theoretically, explicit forms of interactions of magnetic structure of the system such as spatial configuration of magnetic ions in the lattice must be taken into account.
For example, modeling the ESR spectra of intrinsic defects in spin chains is an important problem for which data for finite but rather long chains are necessary~\cite{Sylvain}.

The one dimensional $S=\frac{1}{2}$ XXZ model is one of the most well-investigated systems in the field of ESR study. 
As for the resonance shift and the linewidth of the spectrum which include basic information of the system, a lot of theoretical research has been conducted~\cite{Kanamori-Tachiki,Nagata-Tazuke}. Oshikawa and Affleck~\cite{OA} 
developed an approach based on (1+1)-dimensional field theory, where they used the bosonization method and successfully derived the shift and the linewidth of the resonance peak at low temperatures in the thermodynamic limit~\cite{OA}. This method has been also successfully used to investigate effects of the edge state~\cite{Furuya}.

There are other attempts utilizing the integrability of the XXZ model. Maeda et al.~\cite{maeda} derived the formula for the resonance shift which is exact up to the first order in anisotropy. By applying the Bethe ansatz technique to the formula, they obtained an analytic expression of the resonance shift over all the temperature region. Brockmann et al.~\cite{Weisse} also obtained consistent results focusing on the moments of the spectral shape.
In this way, a lot of information has been found in respect of the resonance shift and the linewidth. 

However, when we need more explicit form of the spectrum which may have a complicated structure, e.g., satellite peaks, long tails, etc., we have to evaluate the Kubo formula for the system Hamiltonian of the interest directly. Such attempts have been also developed~\cite{miyashita,machida,cepas}.

It has been pointed out that the line shape of the XXZ spin chain with even number of spins 
has a double-peak structure at high temperatures for the lattice with finite length $L\le 16$~\cite{cepas}, and the structure has been confirmed up to $L=26$~\cite{ikeuchi}. 
But, its detailed dependence on temperature and size has not been known yet. 
Such information for finite sizes is important to study diluted systems which are ensemble of short chains~\cite{Sylvain},
where temperature-dependence of the spectrum is also important.

In the present paper, 
first, we investigate how the high-temperature spectrum with the double-peak structure at the EPR 
%miya0723
(Electron Paramegnetic Resonance) position,
 i.e., $\hbar\omega_{\rm EPR}=\gamma H$ ($\gamma$ is the gyromagnetic ratio), 
changes to the low temperature spectrum with single peak at a shifted position. 
%miya0723
We present spectra for the  intermediate-temperature region.  
We find a drastic change with the temperature. We also study the case of odd-number of spins, in which
the high-temperature spectrum has a central peak at the EPR position with protuberances beside it.
In this case we also found  the spectrum changes 
to the low-temperature shape which does not depend on the parity of the number of spin.    
The transition is well understood by the energy diagram as a function of the static magnetic field.  

Next, we study origin of the structure of high-temperature spectrum, i.e., the double peak and small peaks beside it (protuberrances).
We find that each peak can be attributed to a  resonance between states with specified magnetizations ($M$ and $M\pm1$).
By making use of this fact, we introduce an extension of the above mentioned moment method~\cite{Weisse} for each contribution from the transition specified by ($M$ and $M\pm1$).
With this analysis, we find that the deviation of 
%miya0724
the mean of the distribution for each contribution to the spectrum reduces as $1/N$.
%ikeuchi0723
This observation strongly 
suggests 
that the structure would shrink to the center in the thermodynamic limit. Indeed the numerical study up to $N=28$ supports
the dependence. 
But on the other hand, we also find that its variance converges to a certain 
finite value as $N$ becomes large. Thus mathematically, there is some possibility of special shape of the spectrum in which the double peak may remain.

%and its variance also reduces as $1/N$.
%This observation strongly suggest that the structure would shrink to the center in the thermodynamic limit, although matematically.....(??). Indeed the numerical study up to $N=28$ support the dependence.
%
The outline of this paper is as follows.
In Sec.~\ref{sec_system}, we introduce the model and 
%miya0724
method. In Sec~\ref{sec_temp_dep}, temperature-dependence of spectra is given. 
In Sec~\ref{sec_decomp}, we analyze structure of spectrum decomposing it into the contributions from transitions between specified sets of magnetizations.
In Sec~\ref{sec_Double}, we study size-dependence of the structure of spectra by applying the moment method to individual contributions from the types of transitions. 
In particular, the estimation of the double peak's separation is an interesting problem, which is discussed in Sec~\ref{sec_separation}.
A summary and discussion of related problems are given in Sec.~\ref{sec_summary}.
%%Appendix A discusses gives the explicit energy structure of small systems.

%%%%%%%%%%%%%%%%%%%%%%%%%%%%%%%%%%%%%%%%%%%%%%%%%%%%%%%%%%%%%%%
%miya0722-2
\section{System and Method}\label{sec_system}

We study a one-dimensional $S=\frac{1}{2}$ XXZ model under a static magnetic field along the $z$-axis. 
We apply to the system
%a perturbative electromagnetic wave with the direction of 
an oscillating magnetic field parallel to the $x$-axis. The total Hamiltonian of the system is given by
\begin{align}
	\mathcal{H_{\mathrm{tot}}}	 	
%&=\mathcal{H}+\lambda(t)\sum_{i=1}^{N}S_{i}^{x}=\mathcal{H}_{0}+\mathcal{H}'+\mathcal{H}_{z}+\lambda(t)\sum_{i=1}^{N}S_{i}^{x}\\
	&=J\sum_{i=1}^{N-1}\bm{S}_{i}\cdot\bm{S}_{i+1}+\Delta\sum_{i=1}^{N-1}S_{i}^{z}S_{i+1}^{z}
	-H\sum_{i=1}^{N}S_{i}^{z}
	+\lambda_{0}\mathrm{cos}\omega t\sum_{i=1}^{N}S_{i}^{x},
	\label{eq:XXZdelta}
\end{align}
where $\Delta$ represents the strength of the anisotropy. In this paper, we set $J=1\mathrm{K}$, $\Delta=-0.08\mathrm{K}$ (i.e., the XY-like anisotropy) and $H=5\mathrm{K}$ (i.e., a sufficiently strong filed), and we impose the open boundary conditions.
%miya0722-2
In the present study we do not include the dipole-dipole interaction, and thus the direction of the chain does not affect the results although the dipole-dipole interaction could cause an interesting angle dependence of the spectrum on the angle between 
the lattice direction and the fields like the Nagata-Tazuke dependence\cite{Nagata-Tazuke}.  
Thus we do not specify it in the present paper.

%miya0724
Here we remind important relations for the the ESR spectrum.
According to the Kubo formula~\cite{kubo_tomita,kubo}, the ESR spectrum, i.e., the absorption rate $I^x(\omega)$ of the oscillating filed can be obtained with the dynamical susceptibility $\chi(\omega)=\chi'(\omega)+\mathrm{i}\chi''(\omega)$ as follows:
\begin{align}
	I^x(\omega)&=\frac{\omega\lambda_{0}^2}{2}\chi''(\omega),\\
	\chi''(\omega)&=\frac{1-\mathrm{e}^{-\beta\omega}}{2}\int_{-\infty}^{\infty}\langle M^x(0)M^{x}(t)\rangle_{\mathrm{eq}}\mathrm{e}^{-\mathrm{i}\omega t}\mathrm{d}t,
\end{align}
where
$M^x(t)=\mathrm{e}^{\mathrm{i}\mathcal{H}t}M^x\mathrm{e}^{-\mathrm{i}\mathcal{H}t}=
\mathrm{e}^{\mathrm{i}\mathcal{H}t}\sum_{i}S_{i}^{x}\mathrm{e}^{-\mathrm{i}\mathcal{H}t}$, $\mathcal{H}=J\sum\bm{S}_{i}\cdot\bm{S}_{i+1}+\Delta\sum S_{i}^{z}S_{i+1}^{z}
	-H\sum S_{i}^{z}$, 
and $\langle\cdots\rangle_{\mathrm{eq}}$ denotes the thermal average with respect to $\mathcal{H}$ at a temperature $\beta^{-1}$.
By using the set of the eigenvalues and eigenvectors $\left\{E_{n},|n\rangle\right\}_{n=1}^{D}$ of the Hamiltonian $\mathcal{H}$ ($D$ is the dimension of the Hilbert space of the Hamiltonian), $\chi''(\omega)$ is readily given by
\begin{align}
	\chi''(\omega)=\sum_{m,n}D_{m,n}\delta(\omega-\omega_{m,n}),
\label{chiomega}
\end{align}
where
\begin{align}
	D_{m,n}\equiv\pi(\mathrm{e}^{-\beta E_{n}}-\mathrm{e}^{-\beta E_{m}})|\langle m|M^x|n\rangle|^2/Z,\quad \omega_{m,n}\equiv E_{m}-E_{n},\quad Z=\sum_{n=1}^D\mathrm{e}^{-\beta E_{n}}.
\label{DD}
\end{align}
Thus, for small systems we obtain $\chi''(\omega)$ by direct numerical estimation of the fomula (\ref{chiomega})
as long as we can obtain all the 
eigenvalues and eigenvectors of the system~\cite{ikeuchi,miyashita}.
%%%0527 
%%greatly 
%used in calculation of the spectrum with the exact diagonalization method~\cite{ikeuchi,miyashita}.
%
%%This model is tractable analytically and a large amount of research has been conducted in various theoretical manners. Besides, much experimental data of materials which are considered as XXZ-like chain are taken.
%%As a short review of previous studies, and also in order to clarify the purpose of our study, we summarize the main results obtained by several analytical methods in Sec.~\ref{Previous_studies}. Then, in Sec.~\ref{Numerical_results} we show our numerical results and discuss the physical analysis on them.
%Estimation of the formula (\ref{chiomega}) by making use of the eigenvalue and eigenvectors of the system obtained exact diagonalization (ED) of the Hamiltonian has been developed to study dependence of spectrum on the lattice structure.\cite{miyashita} 
%%miya0722-1
However, the method is inevitably limited to small systems. Then time-domain methods have been introduced, where the spectrum is obtained by Fourier transform of the autocorrelation function of magnetization (AC method).
There it is known that finite observation time causes artificial modification of the spectrum (say, the Gibbs oscillation). 
Then in our previous study~\cite{ikeuchi}, we proposed a new method (WK method) to make use of the Wiener-Khinchin relation with spectral density of magnetization fluctuation, in which the Gibbs oscillation is suppressed. 
However, in the same time,  we found  that  
that Gibbs oscillation is suppressed in a large system and the AC method works efficiently. 
Thus, in the present work, we obtained  
the spectrum with the AC method.
In the AC method, we can study the double size of the case of the diagonalization theoretically.
However, not the memory but the CPU time prevents us from treating large systems.
In the present paper, we calculated up to $N=28$ in Sec~\ref{sec_Double}.
 The methods are explained in references~\cite{machida,ikeuchi} in detail. 

%by making use of two numerical methods, the autocorrelation (AC) method and the exact diagonalization (ED) method,
%we clarify the temperature-dependence of the ESR spectrum.
%In the AC method, the ESR spectrum is obtained by Fourier transform of the autocorrelation function of the transverse magnetization. This method suffers from the finite observation time (e.g., the Gibbs oscillation) for short chains in general, but 
%it was reported that these effects are smeared out for large systems~\cite{ikeuchi}. The details of the method were explained in the previous paper~\cite{ikeuchi}.

%%%%%%%%%%%%%%%%%%%%%%%%%%%%%%%%%%%%%%%%%%%%%%%%%%%%%%%%%%%%%%%%
\section{temperature-dependence of spectral shapes}\label{sec_temp_dep}

In our previous studies~\cite{cepas,ikeuchi} we studied spectrum at high temperatures, where a double-peak structure was found around the EPR position,
%, i.e., $\hbar\omega_{\rm EPR}=\gamma H$ ($\gamma$ is the gyromagnetic ratio).
On the other hand, the field theoretical study~\cite{OA} gives a single peak with a shift from the EPR position at low temperatures.
Thus, it is interesting to study how the spectrum changes with the temperature.
In this section, we investigate the temperature-dependence of the ESR spectrum. 

\begin{figure}[H]
	%\vspace{-10mm}
	\begin{center}
	\includegraphics[width=59mm]{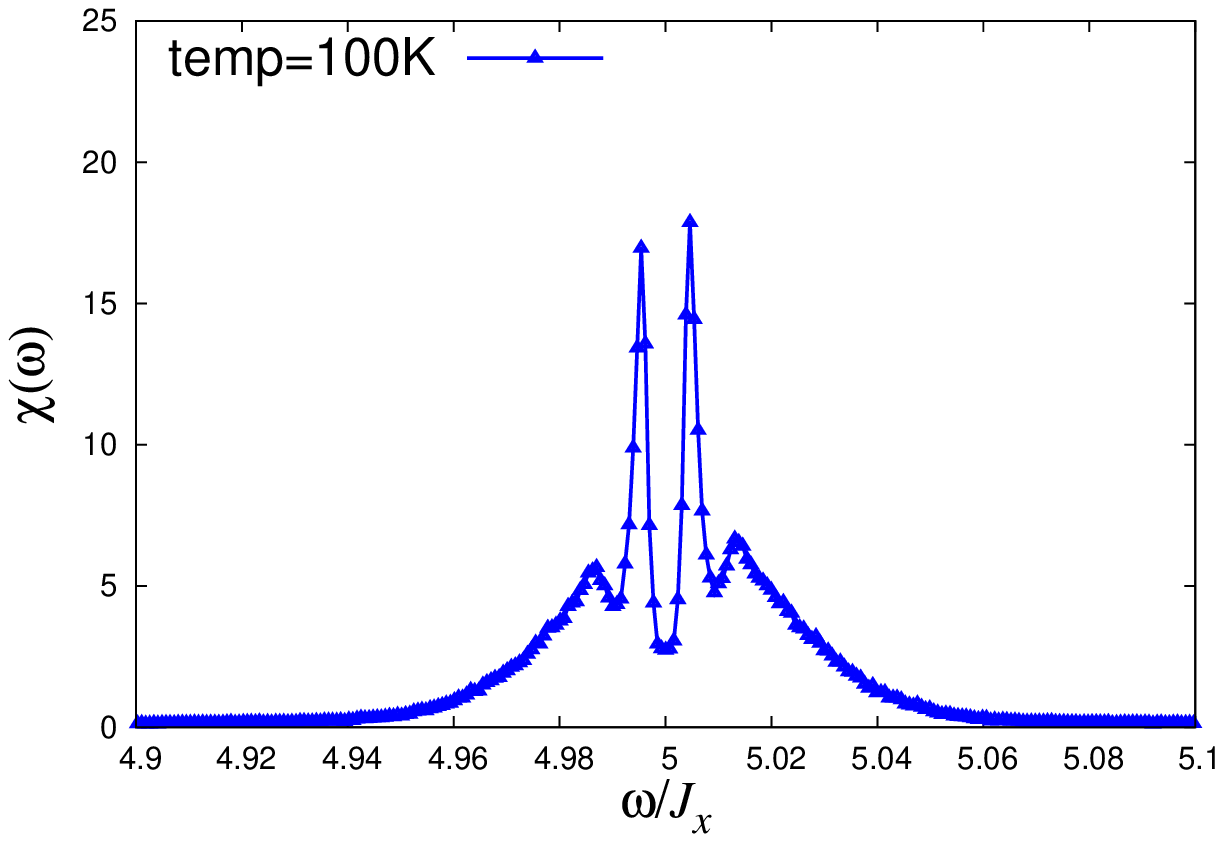}
	\includegraphics[width=59mm]{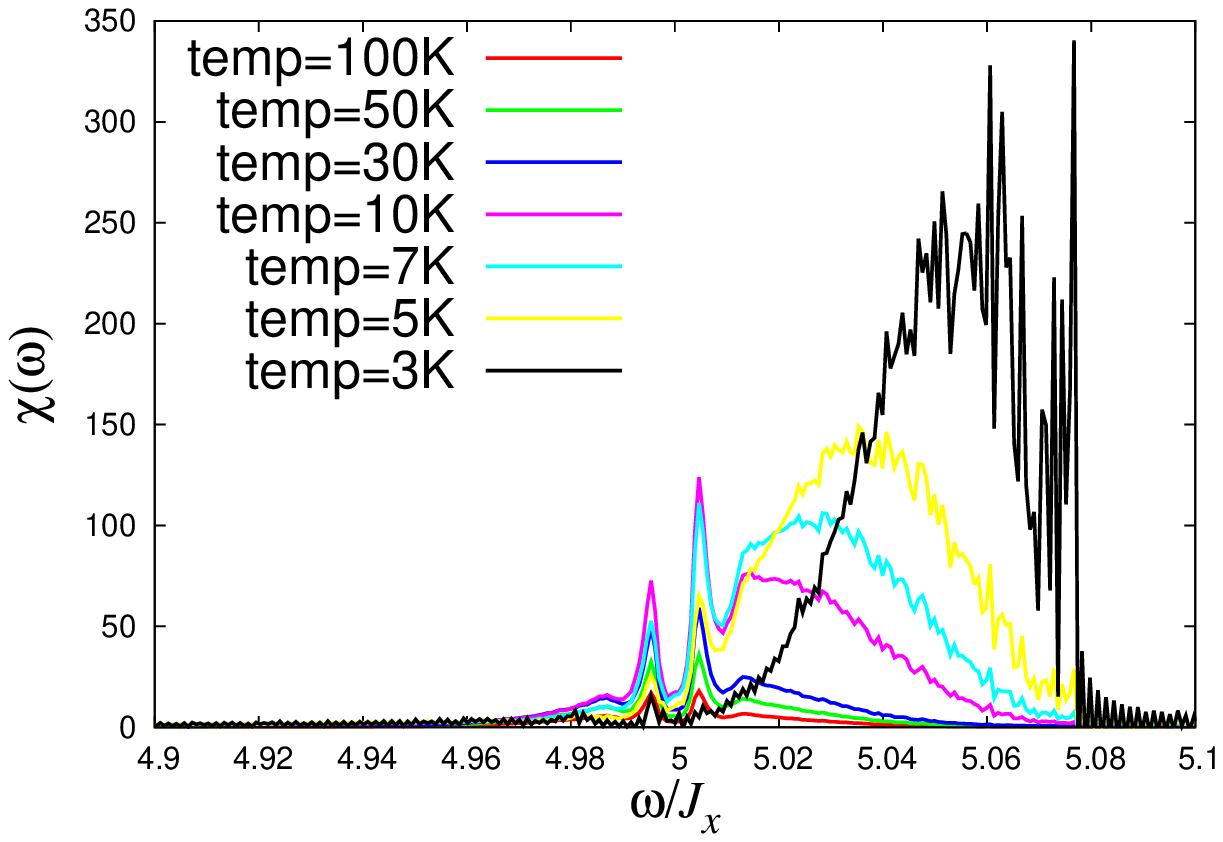}
	\includegraphics[width=59mm]{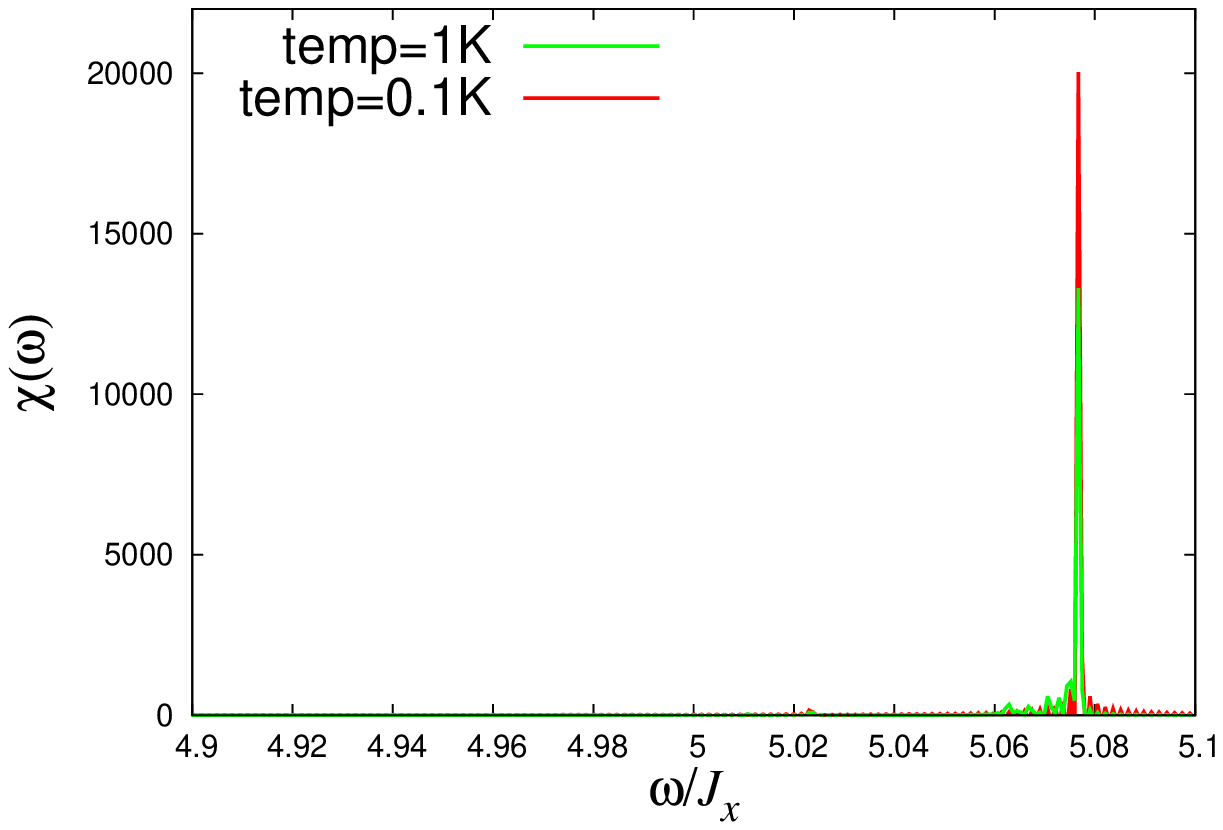}
	\end{center}
	%\vspace{5mm}
\caption{(Color online) (Left) The ESR spectrum for $N=20$ at $\beta^{-1}=100\mathrm{K}$. 
 (Center) The spectra for $N=20$ at $\beta^{-1}=100\mathrm{K}, 50\mathrm{K}, 30\mathrm{K},10\mathrm{K}, 7\mathrm{K}, 5\mathrm{K}$ and $3\mathrm{K}$. 
 (Right) The spectra for $N=20$ at $\beta^{-1}=1\mathrm{K}$ and $0.1\mathrm{K}$ in a 
%%0527
%different 
large scale of $\chi(\omega)$. 
%%0527
%from Left. 
}
\label{temp_dependence_20}
\end{figure}

Let us first discuss the case in which the number of spins is even.
Spectrum for $N=20$ is depicted in Fig.~\ref{temp_dependence_20}(Left) at a high temperature ($\beta^{-1}=100$K), where we find a double-peak structure and two small protuberances next to it.% at high temperatures. 
The center of the structure is located
at the EPR position ($\omega= 5$). 
On the other hand, spectrum at a low temperature ($\beta^{-1}=1$K) has a single peak at a shifted frequency ($\omega\simeq 5.08$), which is depicted in Fig.~\ref{temp_dependence_20}(Right).
Between them, the spectrum shows a significant change with the temperature as depicted in Fig.~\ref{temp_dependence_20}(Center).

\begin{figure}[h]
	%\vspace{-10mm}
	\begin{center}
	\includegraphics[width=60mm]{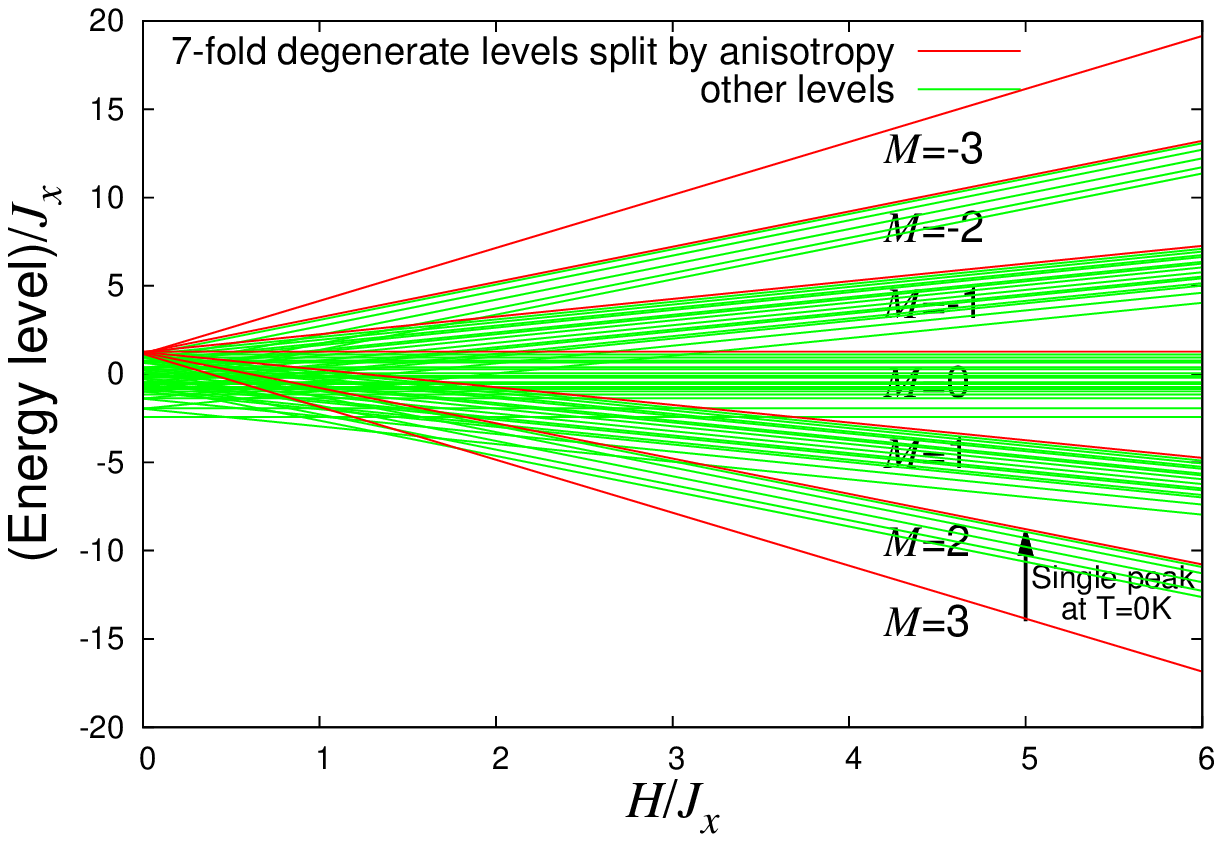}
%	\hspace{1cm}
	%\includegraphics[width=72mm]{level6_mag_kai_3.eps}
 	\includegraphics[width=72mm]{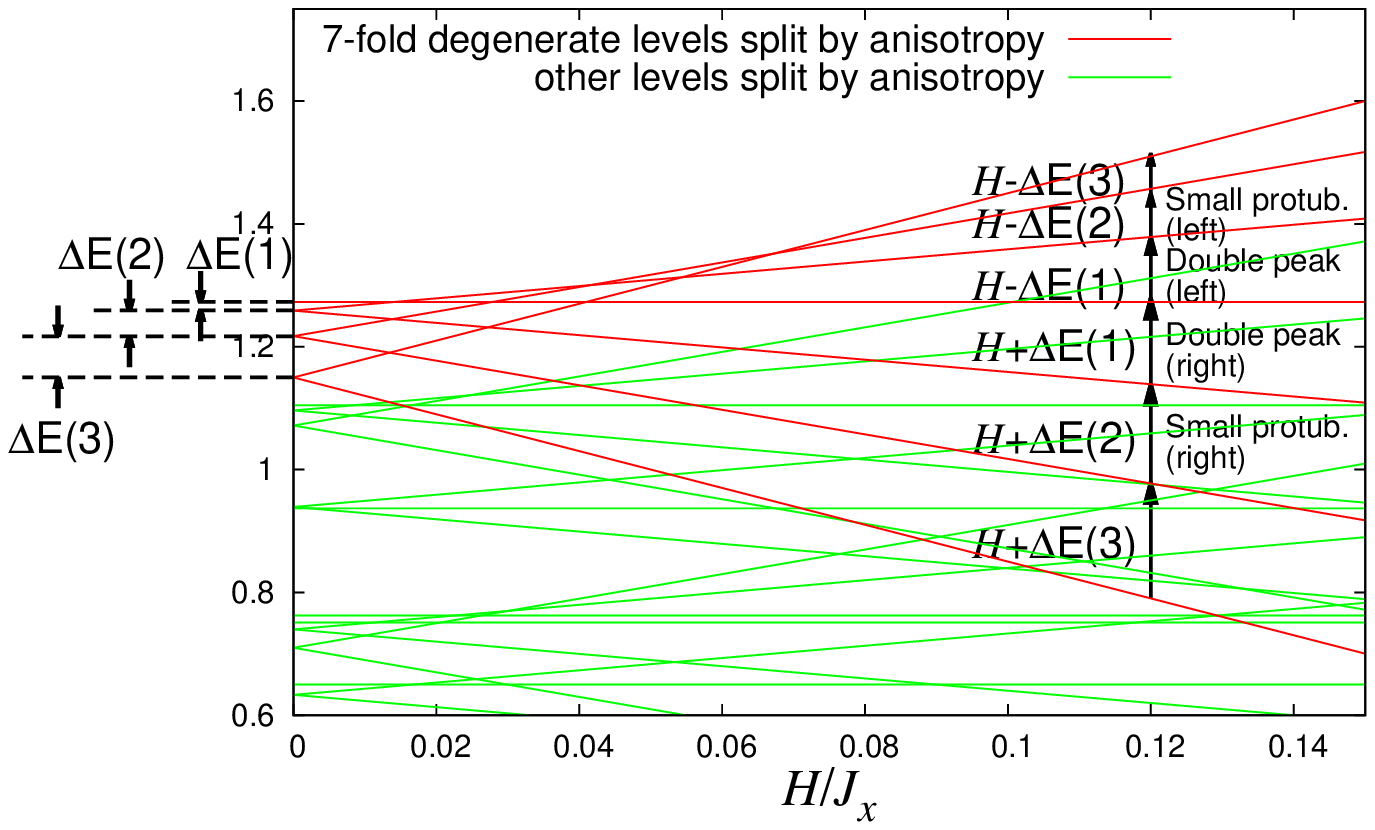}	
	\end{center}
	%\vspace{10mm}
\caption{(Color online) (Left) 64 energy levels for a 6-spin system. We accentuate the 7-fold degenerate levels by the red lines. 
(Right) The enlarged view of the Left figure. 
%When the anisotropy $\Delta=0$, doublets represented by the red lines, i.e., $M=0$, $\left\{M,-M\right\}=\left\{1,-1\right\}$, $\left\{2,-2\right\}$ and $\left\{3,-3\right\}$, come down to 7-fold degenerate states of total magnetization $S=3$ denoted by the blue lines. The energy gap between the doublets with $M=1,-1$ and the single state with $M=0$ is denoted by $\Delta E(1)$. Similarly, $\Delta E(2)$ and $\Delta E(3)$ are defined. 
%The energy split of the state with $M=0$ due to anisotropy is represented by $\delta E_0$.
%From the black arrows representing transitions, we can see that finite $\Delta E(1)$ brings the symmetric double peaks. Here we focused on a multiplet of $S=3$. But these types of transitions occur also at other multiplets though the values of split such as $\Delta E(1)$ differ depending on multiplets. 
%%Note that the values of $\Delta E(1)$'s vary depending on which degenerate levels we focus on.
}
\label{energy_level6}
\end{figure}

In order to understand the change of spectrum, we study the structure of energy diagram as a function of static magnetic field.
The number of states of a system with $N$ spins is $D=2^{N}$. For the large systems, $D$ is too large to draw the diagram in a figure, and thus we draw the diagram for $N=6$ in Fig.~\ref{energy_level6}(Left) (green lines).
There are $2^6=64$ states.
In the case of no anisotropy $\Delta=0$, the eigenstates belong to some multiplets of spin $S$ ($S$ is a positive integer) which has $2(S+1)$-fold degenerate states $(M=-S,-S+1,\cdots, S)$ at $H=0$.
The degenerate states develop with the field according to the Zeeman energy:
$
E(H,M)=E_0-HM,\quad M=-S,-S+1,\cdots, S, %\quad :\quad\text{blue lines in Fig.~\ref{energy_level6} (Right)},
$
where $E_0$ is the energy of the multiplet under no magnetic field.
%%　miya0723
%%図2(右)の青線は消す
%%

%miya0724
Note that there exists only one state that has the magnetization $M=S$, which is 3 in the preset case, 
and the energy is  located at the 
%ikeuchi0723
%highest
lowest
% 
%energy
in the present system as shown in Fig.~\ref{energy_level6}(Left).
%
%When the anisotropy $\Delta=0$, the multiplets of $S=3$ is degenerate at $H=0$. 不要

When the anisotropy is introduced $\Delta\ne 0$, these $2(S+1)$-fold degenerate levels are split into a single state with $M=0$ and $S$ pairs of states which have opposite magnetization, i.e., $\left\{M,-M\right\}=\left\{1,-1\right\}$, $\left\{2,-2\right\}$ and $\left\{3,-3\right\}$ as shown by the red lines.
The energy gap $\Delta E(M)$ between the doublets  with $M=1,-1$ and the single state with $M=0$ at $H=0$ is denoted by $\Delta E(1)$. Similarly, $\Delta E(2)$ and $\Delta E(3)$ are defined. 
%miya0724
Note that $\Delta E(0)=0$ by definition.
Denoting the energy shift of the state with $M=0$ due to anisotropy by $\delta E_0(\ne0)$
%%ikeuchi0723 上と重複
% and 
%the energy gap  between the doublets  with $M$ and the single state with $M=0$ at $H=0$ by $\Delta E(M)$
%%
,
the energy of the state for $\left\{M,-M\right\},\, M=1,2,\cdots S$ is given by
% 
%\beq
%E_0(M)=E_0+\delta E_0+\Delta E(|M|)
%\eeq
\beq
E(H,M)=E_0+\delta E_0+\Delta E(|M|)-HM,\quad M=-S,-S+1,\cdots, S\quad :\quad\text{red lines in Fig.~\ref{energy_level6} (Right)}.
\eeq
%where $\Delta E(M)$ denotes the energy gap between the state with $M$ and the state with $M=0$. 

%, which denotes the energy shift of the state with $M=0$ due to the anisotropy.不要
%%ikeuchi0723
% Thus, the energy levels of the system with anisotropy under the field $H$ are given as
%

%\beq
%E(H)=E_0+\delta E_0+\Delta E(|M|)-HM,\quad M=-S,-S+1,\cdots, S\quad :\quad\text{red lines in Fig.~\ref{energy_level6} (Right)}.
%\eeq
In the left panel, detailed structure due to $\Delta E(M)$ is hardly seen, and thus the magnified structure is given in Fig.~\ref{energy_level6} (Right).

At $H=5$, the ground state is $S=3$ and $M=3$. 
Thus, 
the spectrum at low temperatures is mainly given by the resonance between this state and the excited state with $S=3$ and $M=2$. 
The resonant frequency is changed by $\Delta E(M=3)$ from the EPR value $\gamma H$.
This resonance is the peak in Fig.~\ref{temp_dependence_20}(Right) with the shift $\Delta \omega=\Delta E(M=3)/\hbar$.
On the other hand, at high temperatures 
all the states are 
occupied with nearly the same probability, and the corresponding
spectrum is that in Fig.~\ref{temp_dependence_20}(Left). 
Attribution of each resonance with $\Delta E(M)$ will be investigated in the next section.
At intermediate temperatures, the population distributes with the Boltzmann weight, which gives the change of spectrum 
of Fig.~\ref{temp_dependence_20}.
As the temperature decreases, the double-peak structure breaks down and the spectrum gradually gets shifted to the right side(Center).
%In this scale, only one large peak remains. 不要?

Theoretically temperature-dependence of the line shape has not been studied in detail and so far this drastic change has not been recognized yet. But, this change is robust and it is expected to be observed in corresponding materials.

\begin{figure}[H]
	%\vspace{-10mm}
	\begin{center}
	\includegraphics[width=59mm]{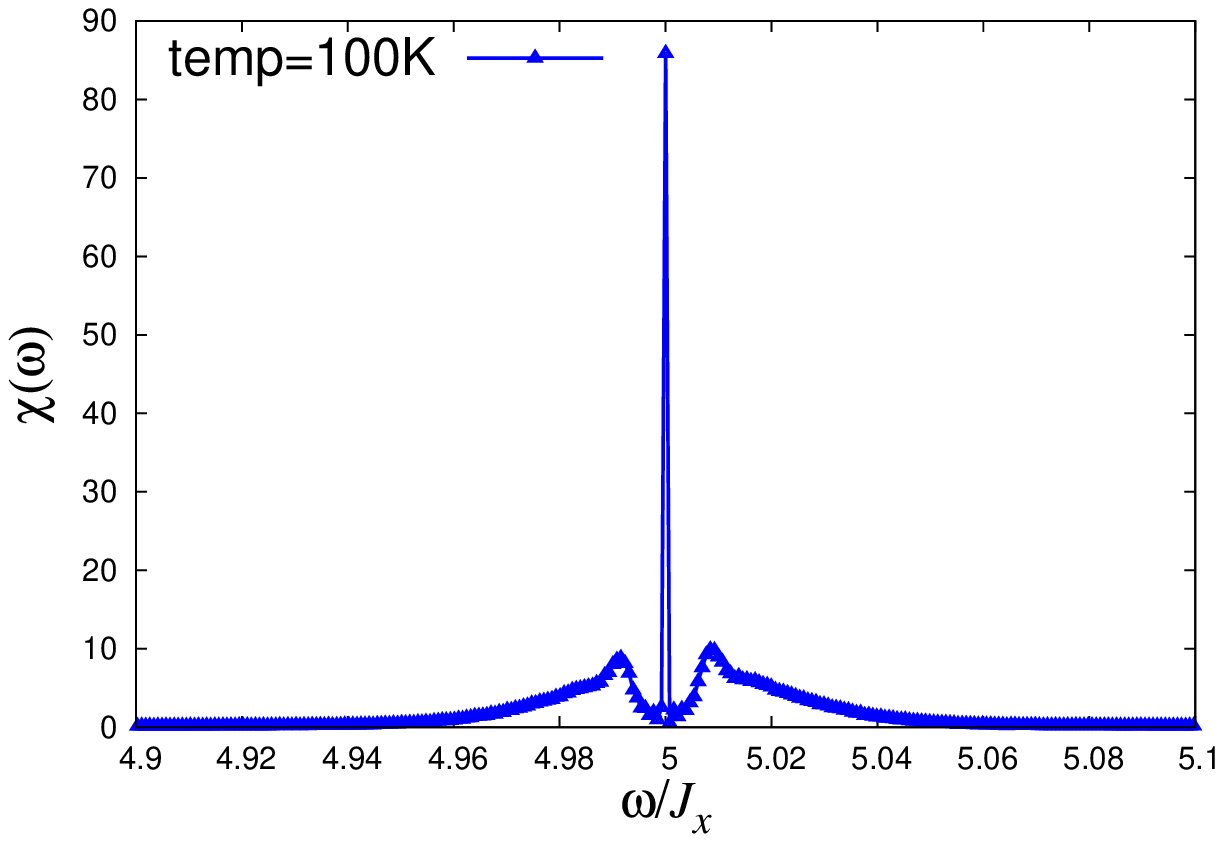}	
	\includegraphics[width=59mm]{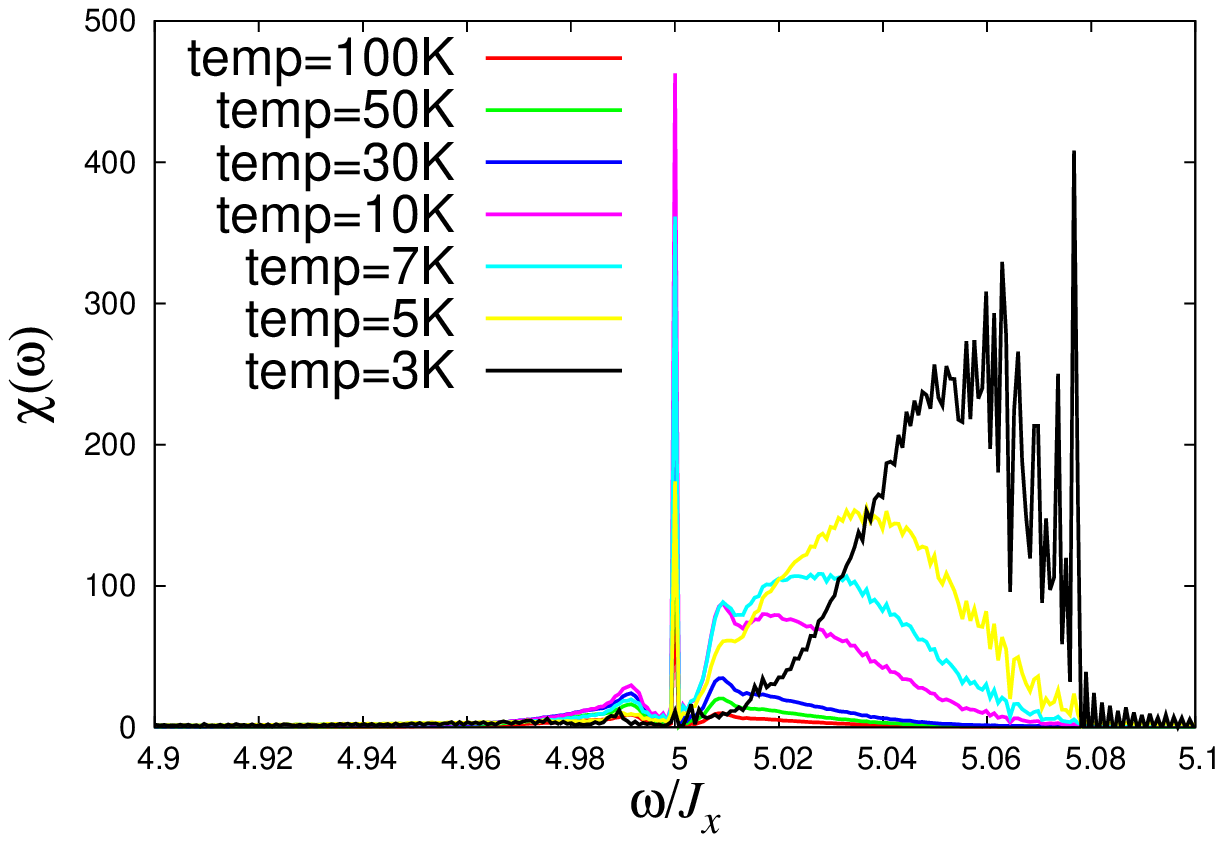}
	\includegraphics[width=59mm]{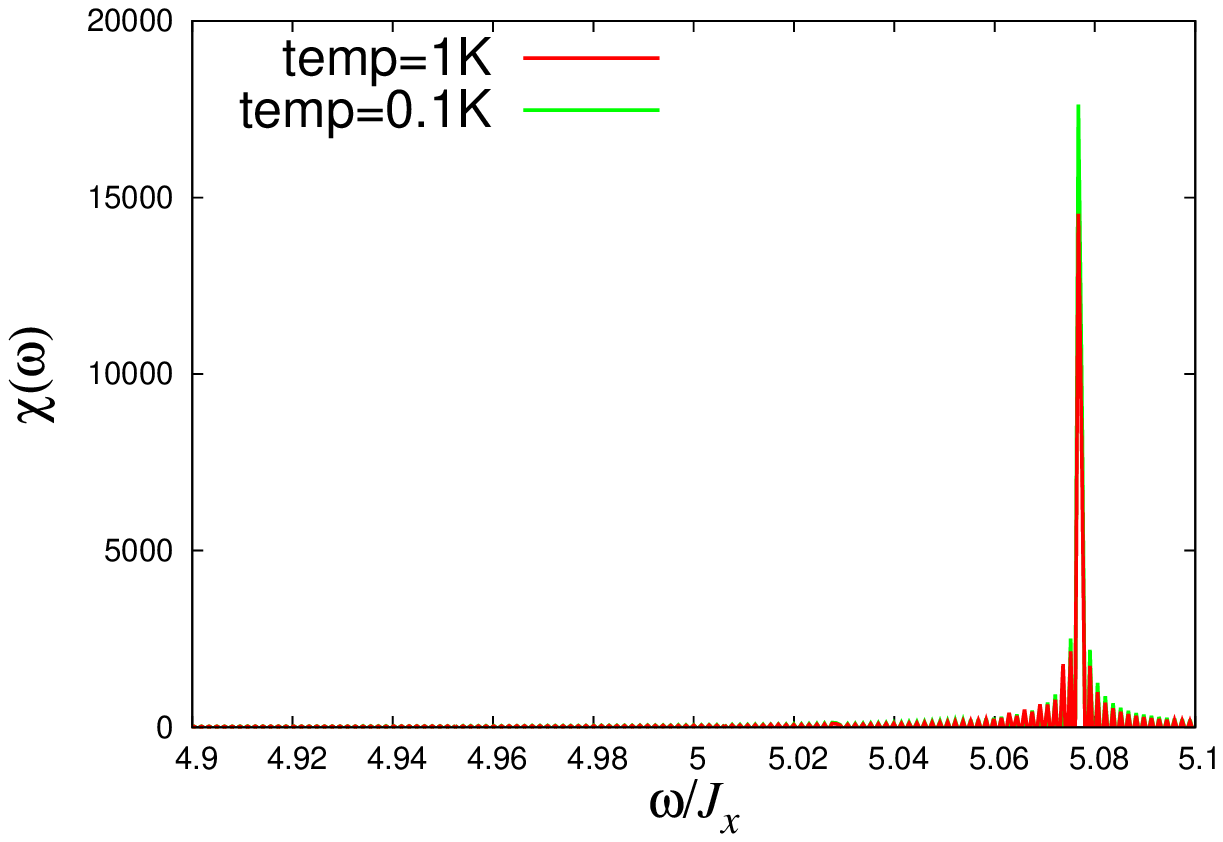}
	\end{center}
	%\vspace{5mm}
	\caption{
(Color online) (Left) The spectrum for $N=21$ at $\beta^{-1}=100\mathrm{K}$.
(Center) The spectra for $N=21$ at $\beta^{-1}=100\mathrm{K}, 50\mathrm{K}, 30\mathrm{K},10\mathrm{K}, 7\mathrm{K}, 5\mathrm{K}$ and $3\mathrm{K}$.  (Right) The spectra for $N=21$ at $\beta^{-1}=1\mathrm{K}$ and $0.1\mathrm{K}$. }
\label{temp_dependence_21}
\end{figure}

%---
Next we study the case of odd number of spins.
In Fig.~\ref{temp_dependence_21} the spectra for $N=21$ are shown.
The spectrum has a single sharp peak at the EPR position with protuberances besides it at the high temperature $\beta^{-1}=100\mathrm{K}$ (Fig.~\ref{temp_dependence_21} (Left)). 
The central peak is a characteristic of the odd number case.
As the temperature is lowered, the spectrum changes to the high frequency side similarly to the case of even number ($N=20$).
 (Fig.~\ref{temp_dependence_21} (Center) and (Right)). 
%miya0724
%The central peak is a characteristic of the odd number case. Its height gets lost and the spectrum with temperature decreasing.

\begin{figure}[h]
	%\vspace{-10mm}
	\begin{center}
	\includegraphics[width=60mm]{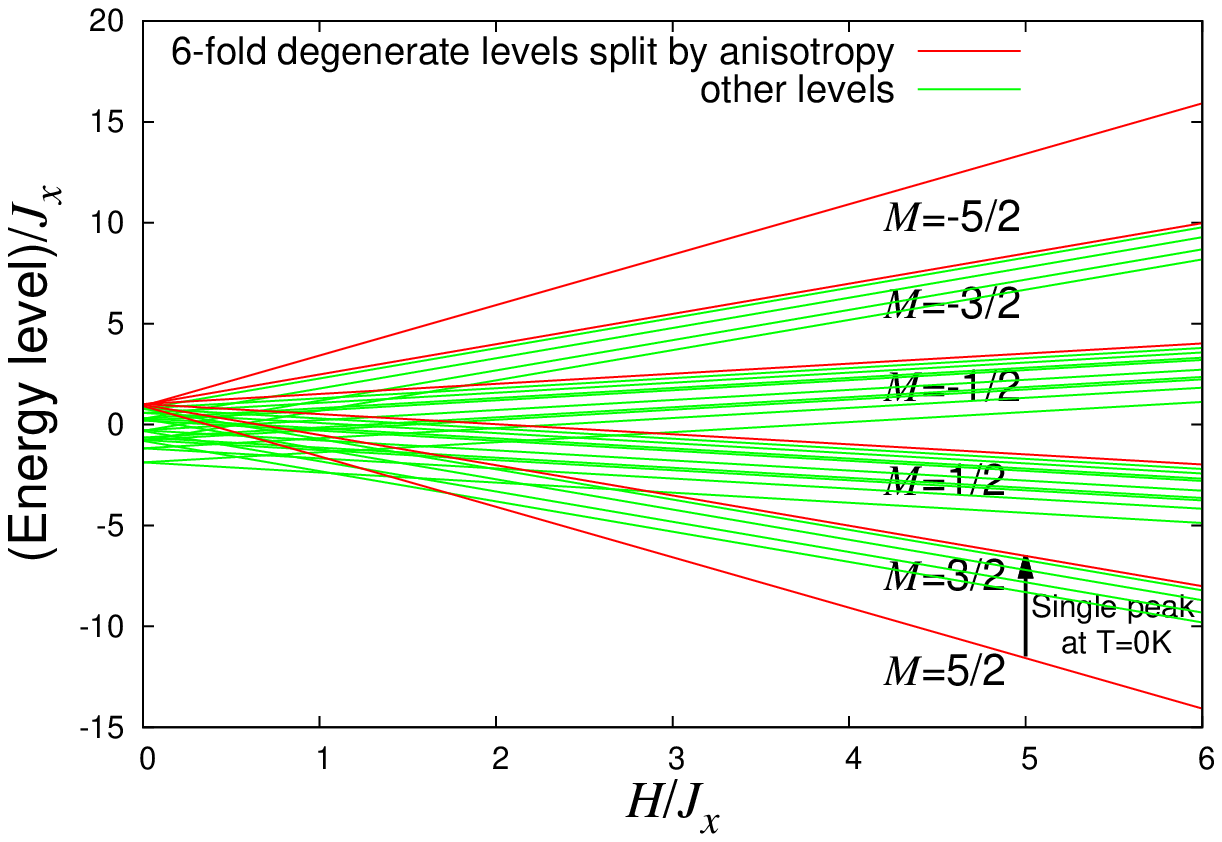}
%	\hspace{-1cm}
	%\includegraphics[width=64mm]{level5_mag_kai_2.eps}
	\includegraphics[width=64mm]{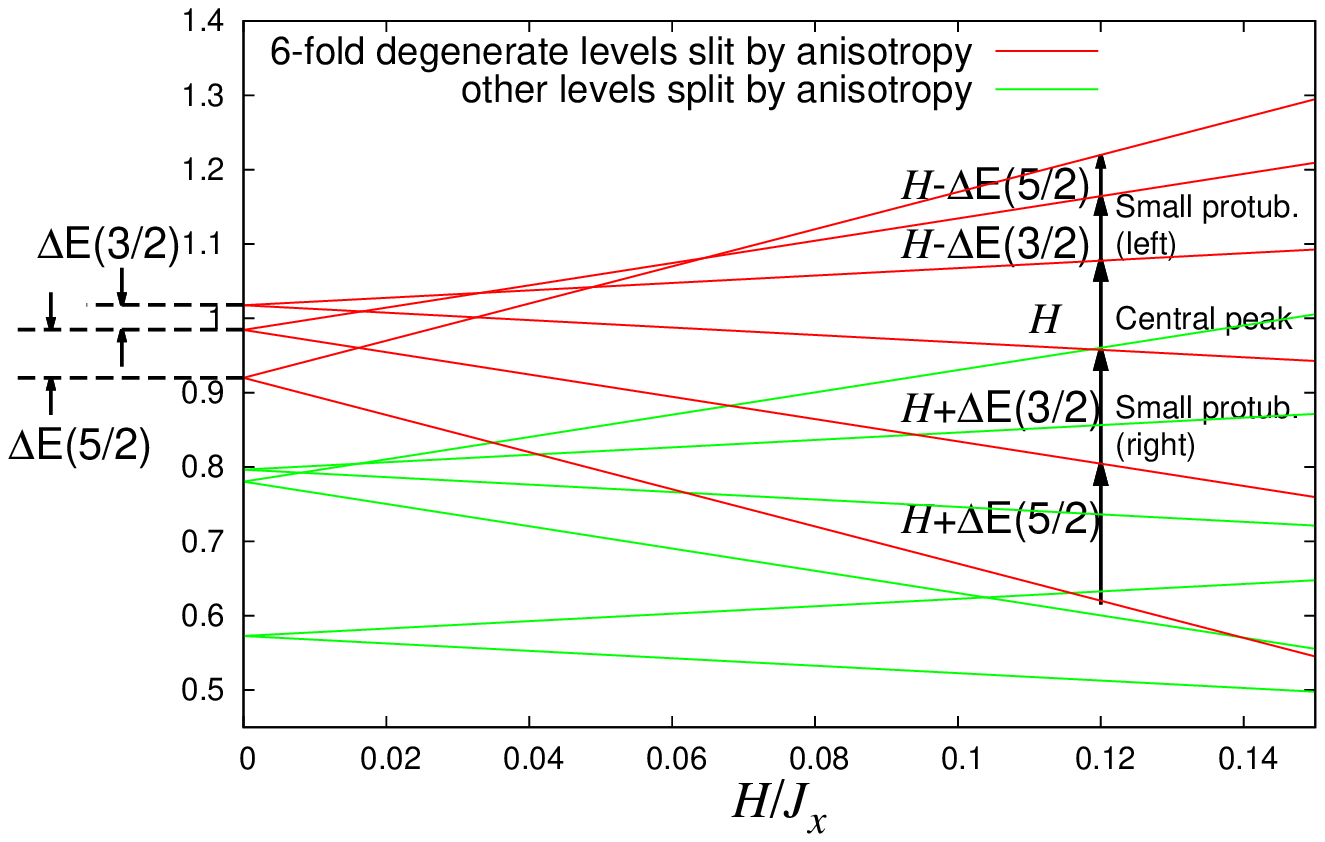}
	\end{center}
	%\vspace{10mm}
\caption{(Color online) (Left) 32 energy levels for a 5-spin system. We accentuate the 6-fold degenerate levels by the red lines. 
 (Right) The enlarged view of the Left figure. }
\label{energy_level5}
\end{figure}

As the corresponding energy diagram we give that for $N=5$  in Fig.~\ref{energy_level5} (Left)
and its magnified one in Fig.~\ref{energy_level5} (Right).
The 6 states %are no more degenerate due to anisotropy and they 
are split into 3 doublets, $\left\{M,-M\right\}=\left\{\frac{1}{2},-\frac{1}{2}\right\}, \left\{\frac{3}{2},-\frac{3}{2}\right\}$ and $\left\{\frac{5}{2},-\frac{5}{2}\right\}$. 
%miya0724
The fact that the state with $M=0$ does not exist is an important difference from the case of even number of spins.
Because of this fact, the spectrum has a sharp peak at EPR position due to the transition between $M=\pm 1/2$.

%The energy gaps between doublets are denoted by $\Delta E\left(\frac{3}{2}\right)$ and $\Delta E\left(\frac{5}{2}\right)$. From the black arrows representing transitions, we can see that finite $\Delta E\left(\frac{3}{2}\right)$ brings the symmetric small protuberances.

%%%%%%%%%%%%%%%%%%%%%%%%%%%%%%%%%%%%%%%%%%%%%%%%%%%%%%%%%%%
\section{Decomposition of the spectrum into contributions from transitions specified by magnetization}\label{sec_decomp}

Without $\Delta$, the spectrum has a single peak at the EPR position, and thus the structure of high temperature spectrum 
should be attributed to the energy structure lifted from the degeneracy.
In what follows, we will reveal the origins of the characteristic shapes observed in the previous section by focusing on the energy diagrams of the systems given in Figs.~\ref{energy_level6} and \ref{energy_level5}

In general, the resonance peaks in ESR spectrum are given by transition between states with $M$ and $M'=M\pm 1$. 
The most dominant contribution comes from the transitions between levels within the same multiplet in the case of $\Delta=0$. The breakdown of SU(2) symmetry due to $\Delta$ allows contributions from transitions between different multiplets, but the contributions from them are found very small. 
Thus, we will ignore those contributions in the analysis of data, although
the contributions are included in spectra obtained by numerical method.
%miya0724 以下の部分は間違い？　無視するのではなくω＜０？？
%(ikeuchi)もともとは、異なるmultiplet間で起こりうる(M→M+1)の遷移について述べていました。
%外部磁場Hが系のエネルギースケールより十分大きいので、（同じmultipletの中ではもちろん）
%異なるmultiplet間でも(M→M+1)の遷移は起こらないor無視できるだろうという意味です。
%\omega<0になるからと言ってもよいと思います。
%
%Besides, we can ignore the contributions from the transitions ($M\to M+1$) because the static filed $H$ is assumed to be sufficiently large compared with the typical energy scale of the system,
%and $e^{-\beta E(M)}\gg e^{-\beta E(M-1)}$. 
%So in the following, we consider only the transitions ($M\to M-1$).
The contributions from the transitions ($M\to M+1$) correspond to emission and it gives spectrum at negative $\omega$.
We do not study those contributions.

Now we decompose the spectrum into contributions from transitions $(M\rightarrow M-1) $of various values of $M$.
Since the multiplet is separated into the pairs of $\{M, -M\}$ and $M=0$,
the contributions of $(M\rightarrow M-1)$ and $(-M+1\rightarrow -M)$ has the same matrix elements.
Thus we classify the contribution according to $M$.
Here we adopt a system with $N=12$ for which there are many states to form a continuous-like lineshape  although we can still calculate the eigenstates by the exact diagonalization method.

\begin{figure}[H]
%\vspace{-10mm}
	\begin{center}
	\includegraphics[width=56mm]{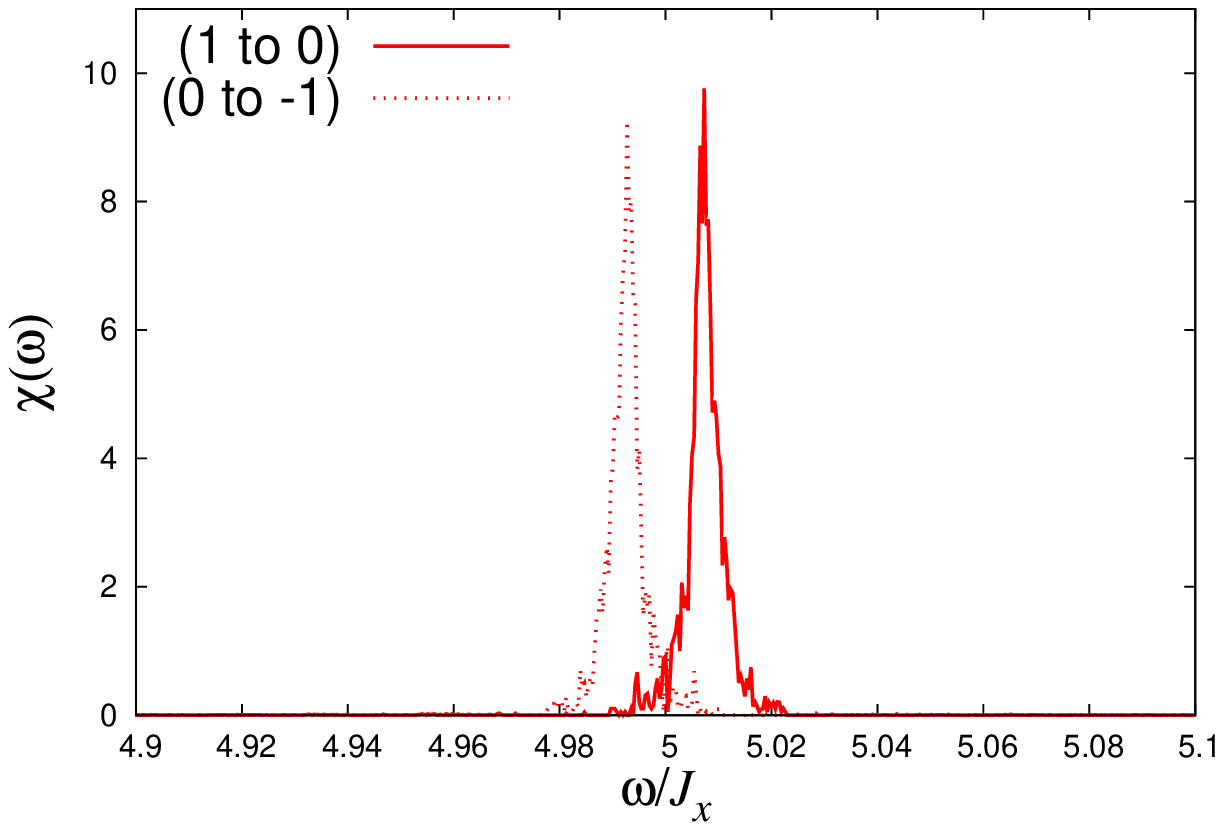}
	\includegraphics[width=56mm]{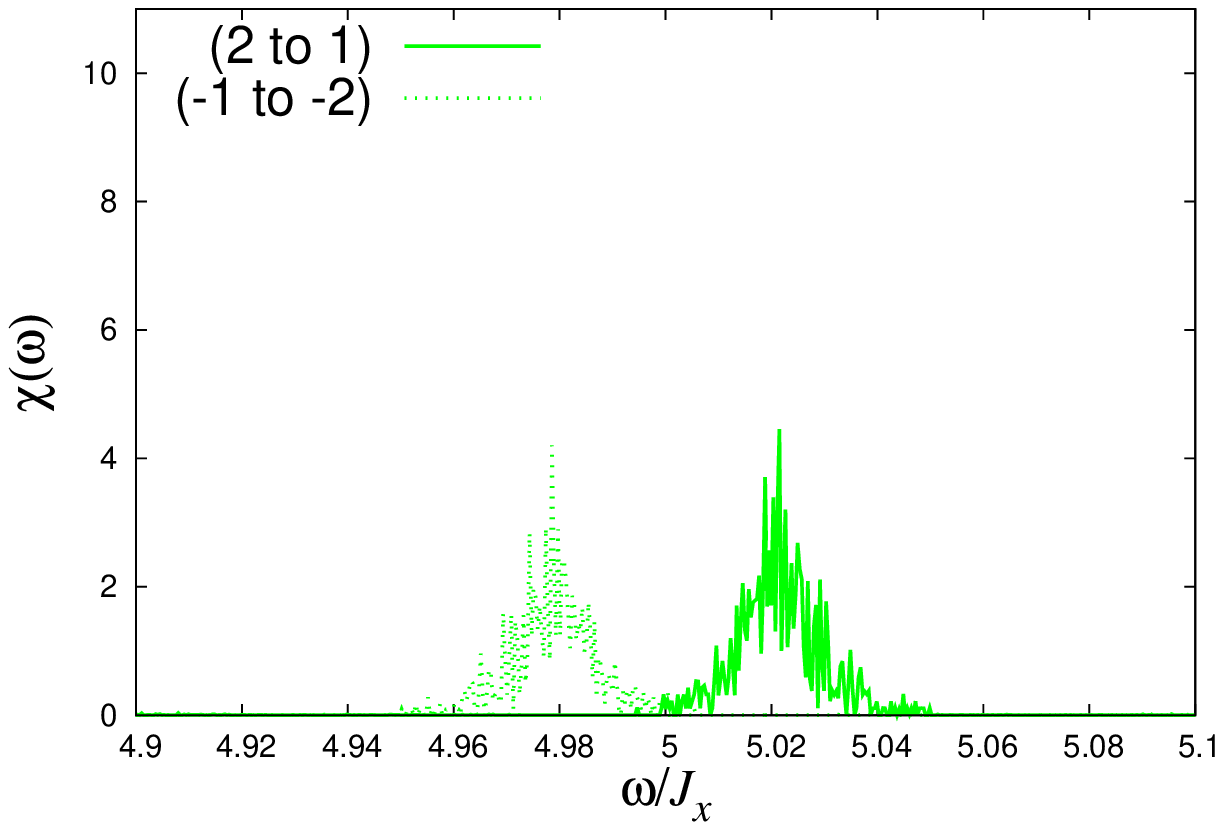}
	\includegraphics[width=56mm]{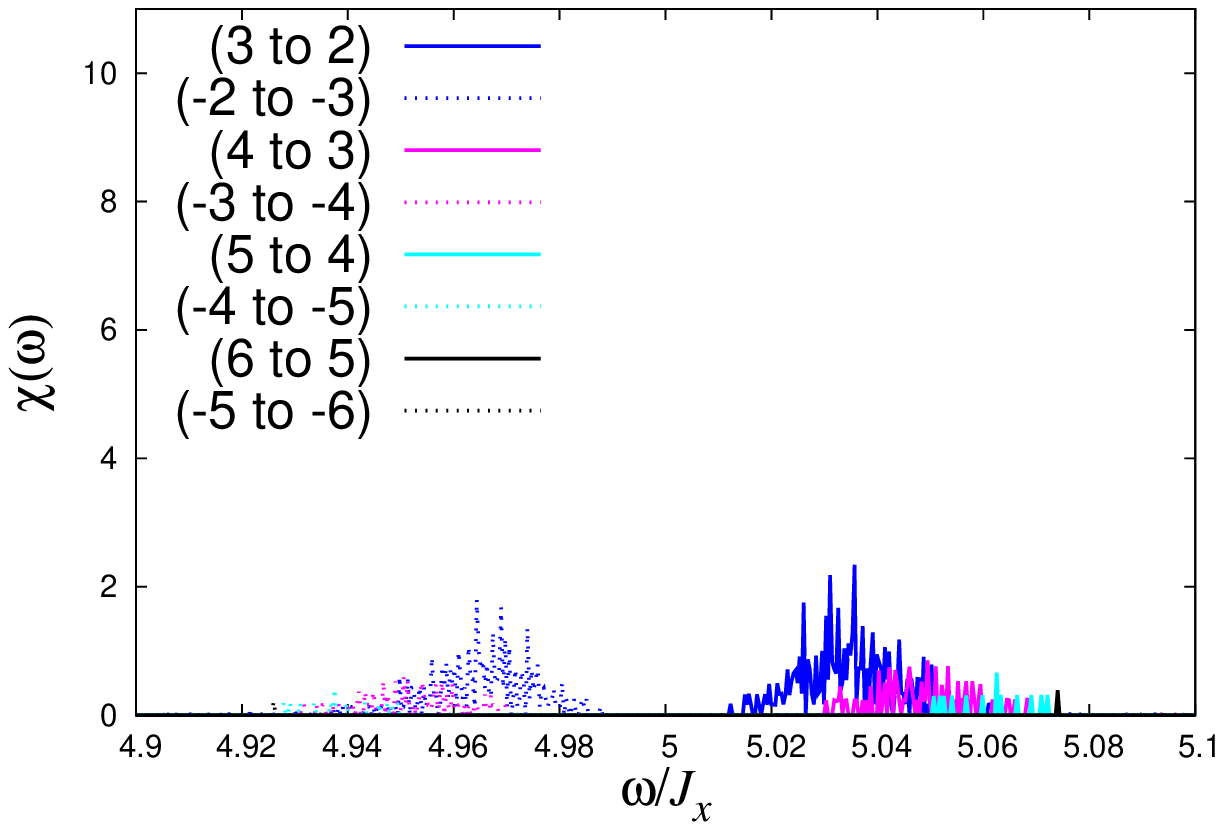}
	\end{center}
	%\vspace{5mm}
\caption{(Color online) Spectra for $N=12$ classified 
%%0527 
by the magnetizations of transitions 
at $\beta^{-1}=100\mathrm{K}$. The Left figure is for the spectrum by transitions $(1\to0)$ (solid) and $(0\to-1)$ (dotted).
The Center figure is for $(2\to1)$ and $(-1\to-2)$. In the Right figure, spectra for larger $M$'s are given:
Blue: $(3\to2)$, $(-2\to-3)$, 
Magenta: $(4\to3)$, $(-3\to-4)$,  
Cyan: $(5\to4)$, $(-4\to-5)$,  
Black: $(6\to5)$, $(-5\to-6)$.}
\label{ED12_color_100}
\end{figure}

In Fig.~\ref{ED12_color_100}(Left), we show the contributions from $M=1$, i.e.,
$(M=1\rightarrow M-1=0)$ (solid line) and 
$(-M+1=0\rightarrow -M=-1)$ (dotted line).
%Here, %in the high-temperature-regime, 
We find the contribution of $(M=1\rightarrow M-1=0)$ gives the right peak of the double peak, and 
that of $(-M+1=0\rightarrow -M=-1)$ the left one.
The separation of the peaks is given by the energy split $\Delta E(M)$.
Here we concern the multiplet with maximum spin $S=N/2=6$, but the transitions of the pair with $M=1$ exist in all the multiplets of $S>0$.  Thus number of transitions between $M=0$ and $\pm 1$ is maximum, and the transition of $M=0$ has the largest contribution. 
The energy gaps $\Delta E(M=0)$ depend on the multiplet which the states belong to, and thus the resonance frequency distributes as we see in Figs.~\ref{ED12_color_100}(Left).

In the same way, the transitions of the pair with $M=2$ give the second peak (protuberance) as depicted in Figs.~\ref{ED12_color_100}(Center), and transitions with $M>2$ make the tail of the spectrum (Figs.~\ref{ED12_color_100}(Right)). 
%More specifically, $(M_1\rightarrow M_1-1)$ gives the right peak of the double peaks, and $(-M_1+1\rightarrow -M_1)$ gives the left one. 

\begin{figure}[H]
%\vspace{-10mm}
	\begin{center}
	\includegraphics[width=56mm]{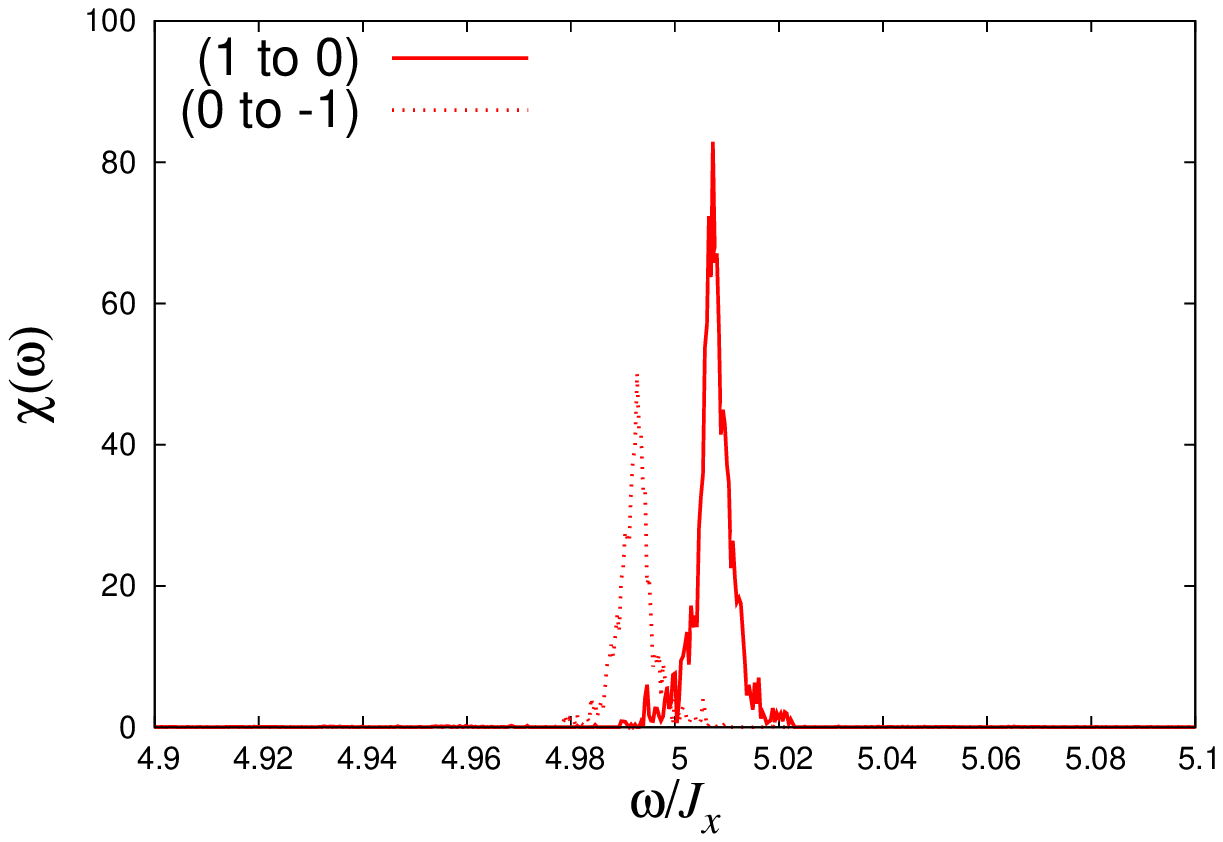}
	\includegraphics[width=56mm]{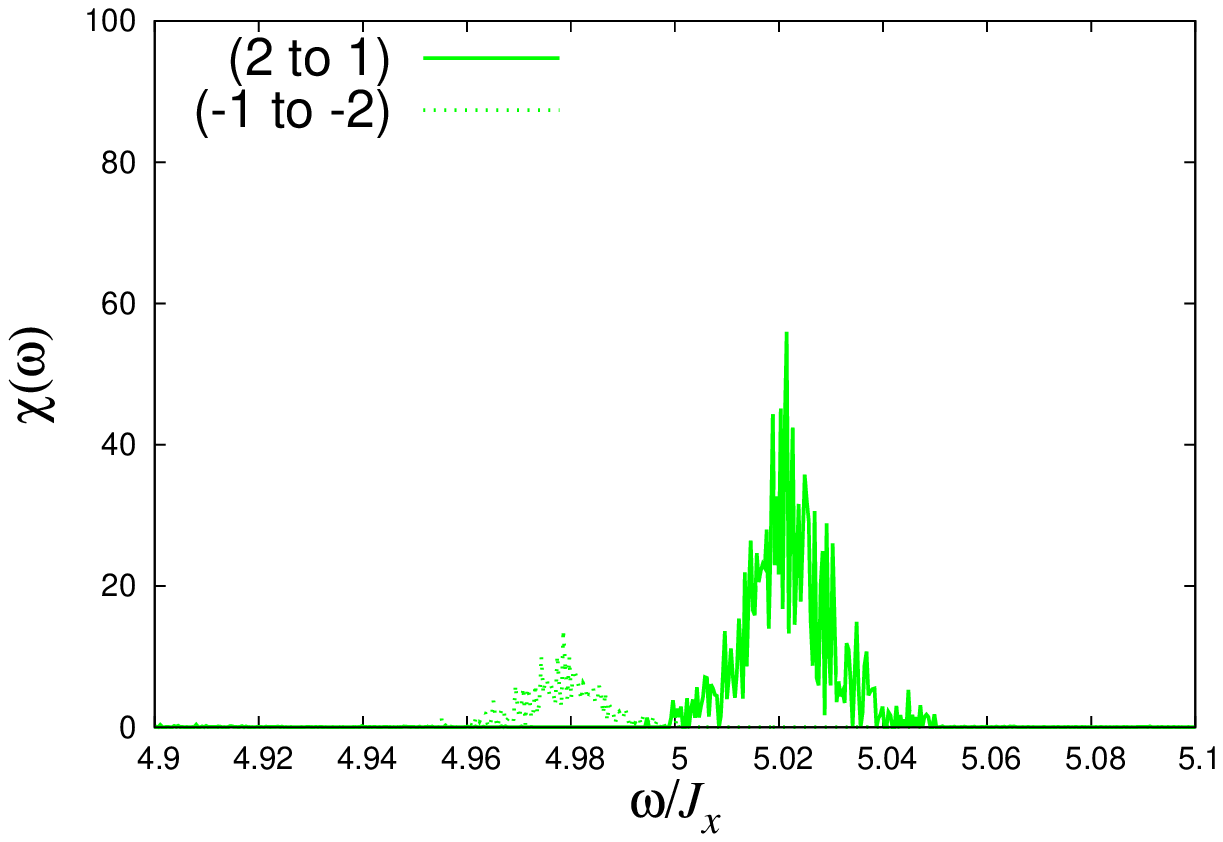}
	\includegraphics[width=56mm]{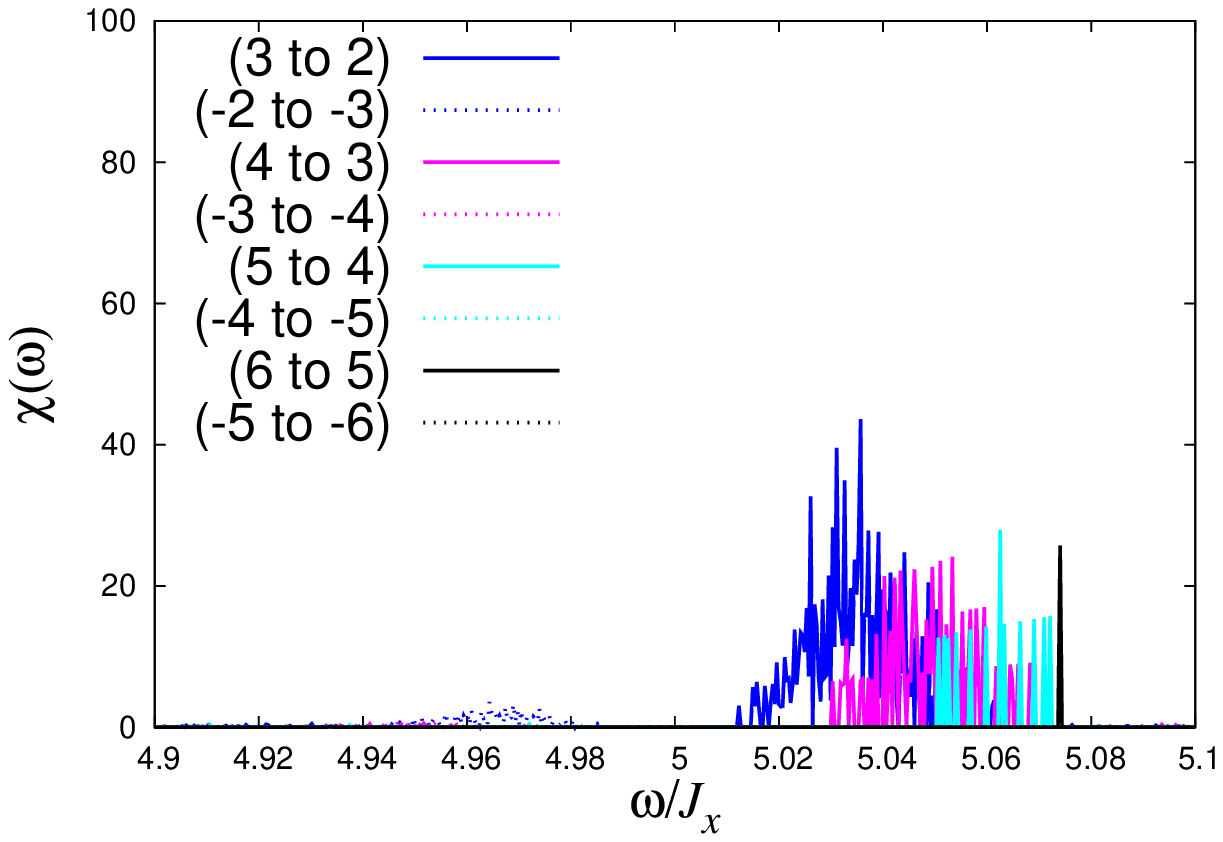}
	\end{center}
	%\vspace{5mm}
\caption{(Color online) Spectra for $N=12$ classified 
%%0527 
by the magnetizations of transitions 
at $\beta^{-1}=10\mathrm{K}$. The Left figure is for the spectrum by transitions $(1\to0)$ (solid) and $(0\to-1)$ (dotted).
The Center figure is for $(2\to1)$ and $(-1\to-2)$. In the Right figure, spectra for larger $M$'s are given: 
Blue: $(3\to2)$, $(-2\to-3)$, 
Magenta: $(4\to3)$, $(-3\to-4)$,  
Cyan: $(5\to4)$, $(-4\to-5)$,  
Black: $(6\to5)$, $(-5\to-6)$.}
\label{ED12_color_10}
\end{figure}

In the intermediate-temperature regime(Figs.~\ref{ED12_color_10}), the Boltzmann factors 
%$\exp(-\beta E_0(m)+\beta Hm)$
$\exp(-\beta E(H,m))$ for m=$M$ and $-M+1$ are gradually getting difference,  and thus
the spectral shape becomes no more symmetric around $\omega_{\rm EPR}$, 
and consequently the transitions between levels with large $m=M>0$ become larger in the spectrum
studied in the previous section. 
This explains the temperature-dependence of the spectrum. 

\begin{figure}[H]
%\vspace{-10mm}
	\begin{center}
	\includegraphics[width=56mm]{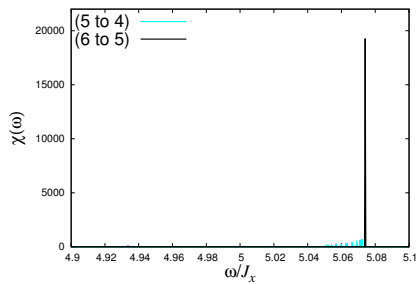}
	\end{center}
	%\vspace{5mm}
\caption{(Color online) Spectra for $N=12$ classified 
%%0527 
by the magnetizations of transitions 
at $\beta^{-1}=1\mathrm{K}$. 
This figure is depicted according to the same notation as in Figs.~\ref{ED12_color_100} and \ref{ED12_color_10}. But the black solid peak corresponding to the transition $(6\to5)$ mainly contributes to the spectrum, and the peaks from transition between smaller $M$'s can hardly be observed in this temperature range.}
\label{ED12_color_1}
\end{figure}

In the low-temperature regime (Figs.~\ref{ED12_color_1}), the state of 
$M=S$ 
is mostly populated under a large filed, and the transition from it to $M=S-1$ gives the single peak with a shift. 

\vspace{-3mm}

For odd number of spins for which we depicted the energy structure in Fig \ref{energy_level5}, the multiplets consist of half-odd spins $S$ of $M=-S,-S+1,\cdots, -1/2,+1/2,\cdots, S$. The transitions between $M=1/2$ and $-1/2$ are most populated and they give the central peak. As the temperature increases, very similarly to the even number case, the peaks shift to the high frequency side, and finally become a single sharp peak with a shift.
%%
%%miya0723 図８は割愛
%%

% (Fig.~\ref{ED11_color}).
%
%%
%%
%%
%\begin{figure}[H]
%	%\vspace{-10mm}
%	\begin{center}
%	\includegraphics[width=56mm]{N11_temp100_trans_black.eps}
%	\includegraphics[width=56mm]{N11_temp10_trans_black.eps}
%	\includegraphics[width=56mm]{N11_temp1_trans_black.eps}
%	\end{center}
%	%\vspace{10mm}
%\caption{(Color online) Spectra for $N=11$ classified by using 6 different colors. The red line denotes the spectrum by transitions $(\frac{1}{2}\to-\frac{1}{2})$. Similarly, each color corresponds to a different type of transitions as follows. 
%Green: $(\frac{3}{2}\to\frac{1}{2})$, $(-\frac{1}{2}\to-\frac{3}{2})$,  
%Blue: $(\frac{5}{2}\to\frac{3}{2})$, $(-\frac{3}{2}\to-\frac{5}{2})$,  
%Magenta: $(\frac{7}{2}\to\frac{5}{2})$, $(-\frac{5}{2}\to-\frac{7}{2})$,  
%Cyan: $(\frac{9}{2}\to\frac{7}{2})$, $(-\frac{7}{2}\to-\frac{9}{2})$,  
%Black: $(\frac{11}{2}\to\frac{9}{2})$, $(-\frac{9}{2}\to-\frac{11}{2})$. (Left) $\beta^{-1}=100\mathrm{K}$. (Center) $\beta^{-1}=10\mathrm{K}$. (Right) $\beta^{-1}=1\mathrm{K}$.}
%\label{ED11_color}
%\end{figure}	
%%

%%%%%%%%%%%%%%%%%%%%%%%%%%%%%%%%%%%%%%%%%%%%%%%%%%%%%%%%%%%%%%%
%
%miya0724 章にする。
%

\section{Size-dependence of spectral shapes}\label{sec_Double}

\subsection{Numerical observation}

\begin{figure}[H]
	%\vspace{-10mm}
	\begin{center}
	\includegraphics[width=70mm]{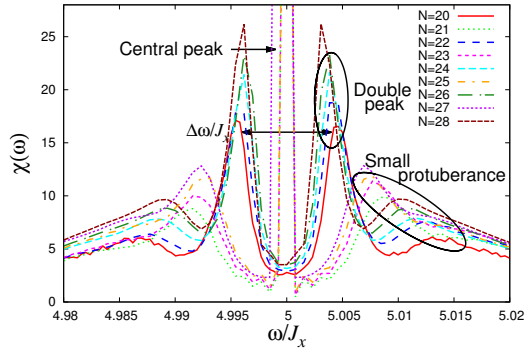}	
	\end{center}
	%\vspace{10mm}
\caption{(Color online) Spectra of $N=20,21,22,23,24,25,26,27$ and $28$ depicted on the unified scale. 
Note that the central peaks of odd systems are cut off in this figure. }
\label{even_odd}
\end{figure}

Now, we examine the size-dependence of characteristic shapes of spectra found in the high-temperature region, such as the double-peak structure.
%%ikeuchi0723 必要か?
% Here we investigate only the case of even number of spins.
%
%First let us show the numerical results obtained by the AC method. 
%Since the double-peak structure is found at a high temperature, we set $\beta^{-1}=100\mathrm{K}$ in the simulation. 
We depicted spectra of various sizes $N$ in Fig.~\ref{even_odd}, where we see rather systematic size-dependence.
The systems with even numbers of spins show that
the separation between double peaks $\Delta\omega$ decreases as the system size becomes large. 
We may anticipate that in the thermodynamic limit, the double peak may get stuck together to become a single central peak. 
In fact, this problem has already posed in the reference~\cite{cepas}, but its answer has not been concluded yet. Besides, the separation between two small protuberances also seems to decrease with the system size increased, and shrink to the center in the thermodynamic limit.
%Both systems with even and odd numbers of spins show that the protuberances also shrink to the center.
%Quantitative dependence of the size will be given later.

\begin{figure}[H]
	\begin{center}
	\includegraphics[width=56mm]{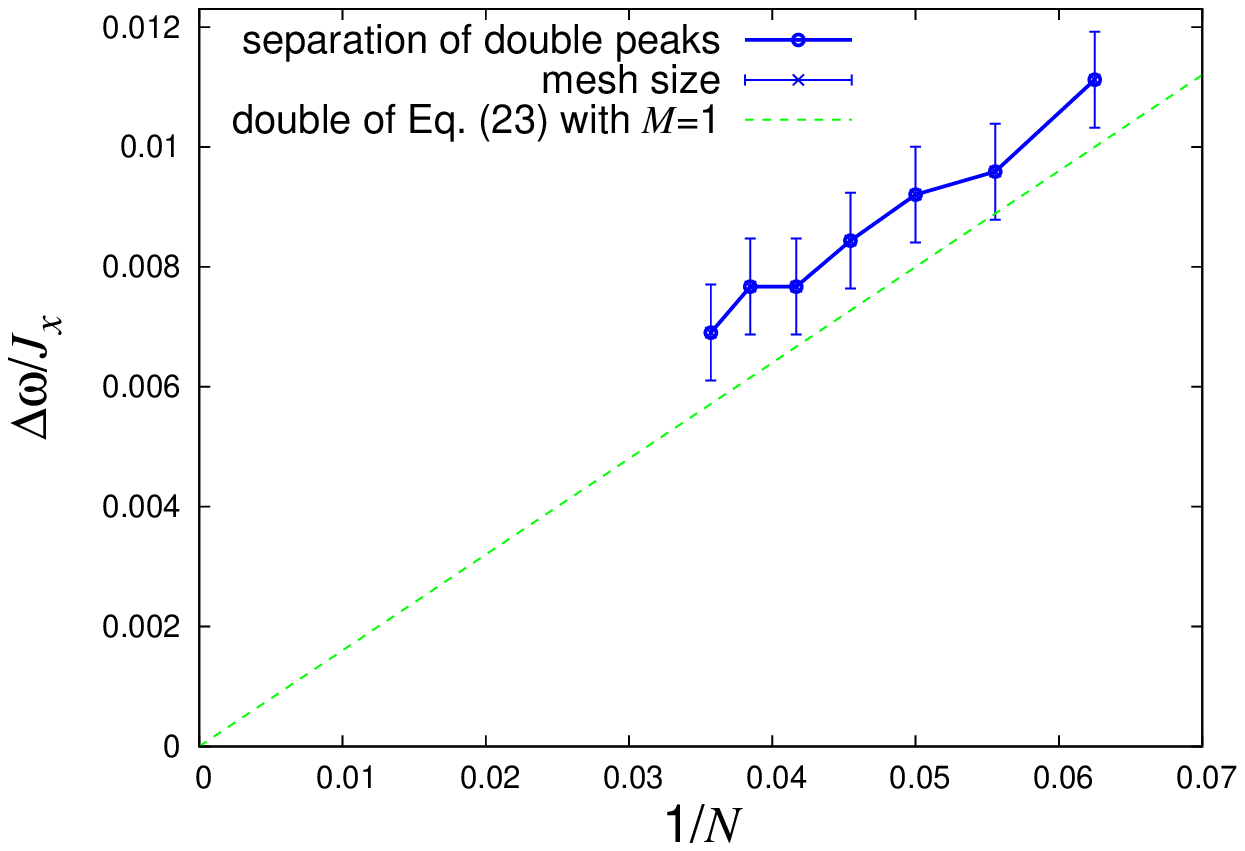}
	\includegraphics[width=56mm]{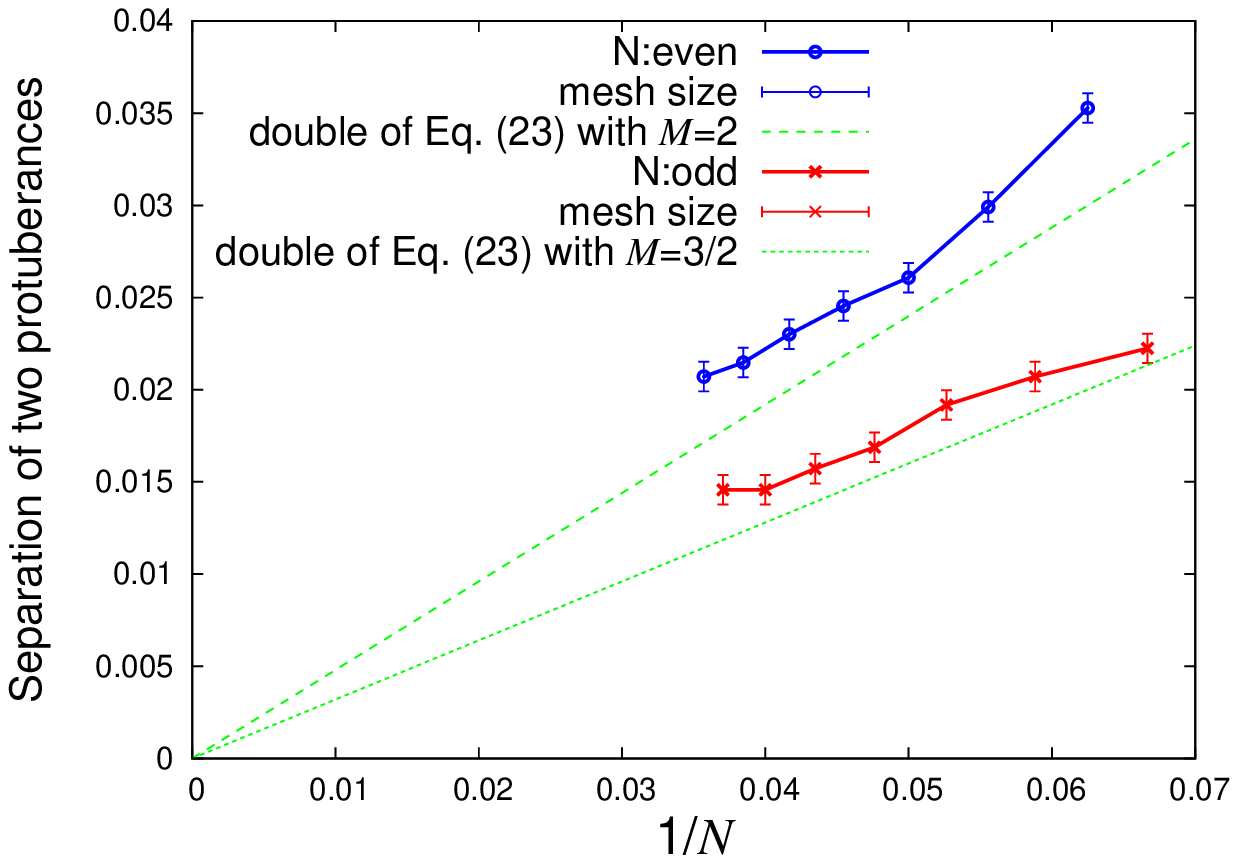}
	\includegraphics[width=56mm]{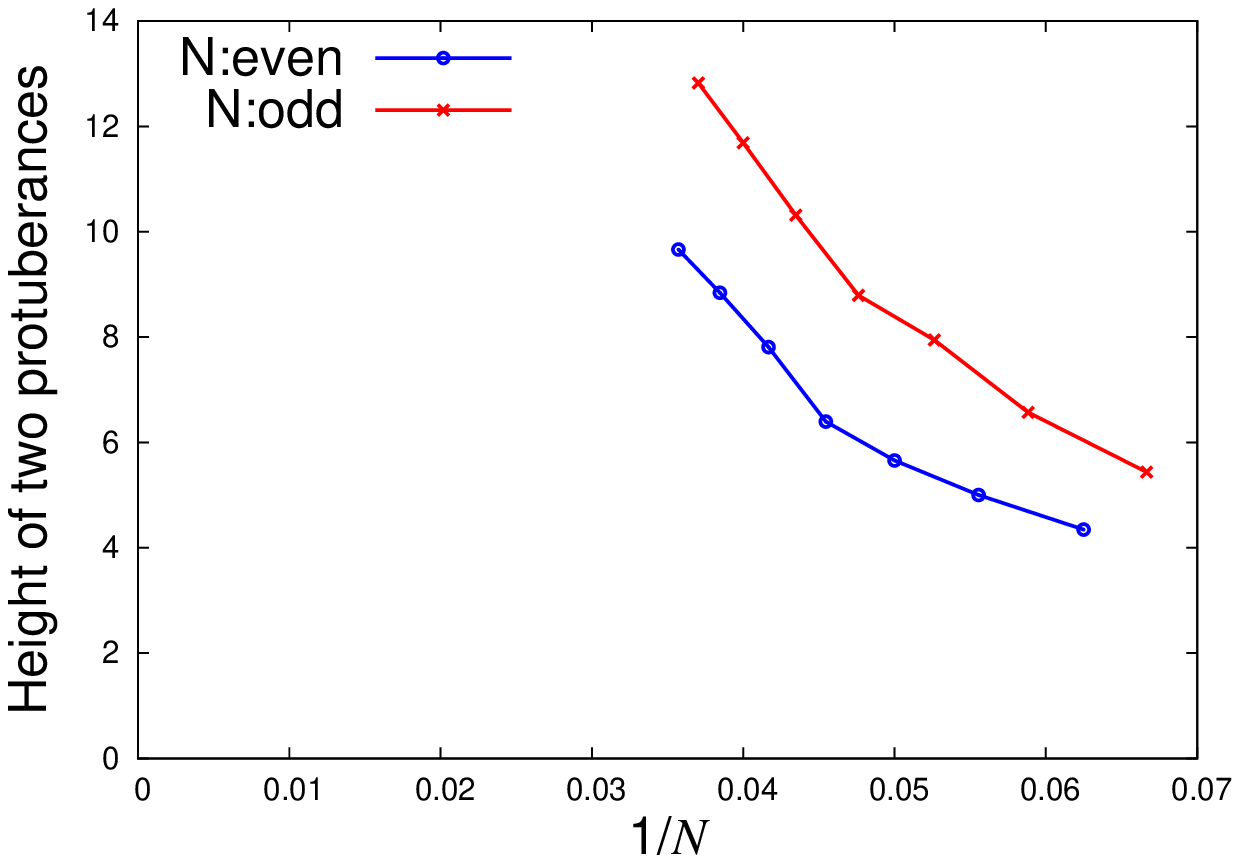}
	\end{center}
	%\vspace{10mm}
\caption{
(Color online) (Left) Size-dependence of the separation of the double peak. There seems to be a tendency for separation of the double peak to decrease as $N$ becomes large. But it might be saturated at some point. The green line denotes theoretical values calculated by Eq.~(23).
(Center) Size-dependences of the separation of two small protuberances. The red points denote odd-spin systems and the blue points even-spin systems. We can find that the small protuberances come close to the center $\omega=H$ as the system size $N$ increases. The error bars denote the mesh size of $\Delta\omega$ determined by the observation time $T$.
The green lines denote theoretical values calculated by Eq.~(23). 
(Right) Size-dependences of the height of two small protuberances. The height is related to the the intensity of the absorption. The protuberances become larger with $N$ increasing.}
	\label{size_dependence}
\end{figure}

The size-dependence 
of the separation $\Delta\omega$ between the double peak is given in Fig.~\ref{size_dependence} (Left). The error bars in the figure denote the mesh size of $\Delta\omega$ given by the observation time $T$ as $2\pi/T$. Here we find that in even systems the separation roughly decreases with the size as $1/N$ at least up to $N=22$, and that the separations of $N=22, 24$ and $26$ are almost the same. But at $N=28$ it again decreases. 
Thus, although we can have a rough picture about the finite size effect, we cannot conclude how the separation behaves in the large-size limit from the figure.
In Fig.~\ref{size_dependence} (Center) we plot the size-dependence of separations of two small protuberances of both even and odd cases.
This shows that the separation of protuberances decreases roughly proportionally to $1/N$. 
% which also show decrease roughly proportionally to $1/N$. 
In Fig.~\ref{size_dependence} (Right), we find the heights of the peaks of protuberances increase with the system size.

\subsection{Estimation of the separation}\label{sec_separation}

Now, we examine the size-dependence of characteristic shapes of spectra by making use of the moment method.
The method of moments originated in van Vleck's paper~\cite{vanvleck} and has been used as a basic tool to investigate the spectral shapes~\cite{kubo_tomita,anderson}. In a recent study~\cite{Weisse}, the moments of the whole spectrum for the XXZ chain were discussed in detail.
The width and position of a peak in spectrum have been studied by the the moment method. 
%In what follows, we aim to understand the above numerical results analytically by the moment method. 
In the moment method, we can grasp characteristics of ESR spectra by focusing on $n$-th order moment 
$m_{n}$ of the spectrum. 

Since we are considering
properties at high temperatures,
we use $S_{xx}(\omega)$ instead of the susceptibility $\chi(\omega)$:
\begin{align}
S_{xx}(\omega)=\int_{-\infty}^{\infty}\langle M^xM^x(t)\rangle_{\mathrm{eq}}\mathrm{e}^{-\mathrm{i}\omega t}\mathrm{d}t.
\end{align}
The moment $m_{n}$ is defined as\cite{Weisse}:
\begin{align}
m_{n}&=\int_{-\infty}^{\infty}\mathrm{d}\omega\omega^n
\int_{-\infty}^{\infty}\mathrm{d}t\langle M^xM^x(t)\rangle_{\mathrm{eq}}
	\mathrm{e}^{-\mathrm{i}\omega t}\\
	&=2\pi\langle M^x(\mathrm{ad}_{\mathcal{H}})^nM^x\rangle_{\mathrm{eq}},
\end{align}
where $\mathrm{ad}_{\mathcal{H}}\cdot\equiv\left[\mathcal{H},\cdot\right]$. 
We could estimate these quantities by exact solutions or numerical methods. 
In particular, 
at the infinite temperature, we obtain the moments from combinatorics
as we see later. 

The intensity (i.e., the area of the spectrum), the mean position, and the linewidth are defined as
\begin{align}
&\mbox{intensity:}\quad m_{0},\\
&\mbox{mean position:}\quad \frac{m_1}{m_0},\\
&\mbox{linewidth:}\quad\sqrt{\frac{m_2}{m_0}-\left(\frac{m_1}{m_0}\right)^2},
\label{linewidth}
\end{align}
respectively. 
There is still room for argument about the definition of linewidth given in (\ref{linewidth}). Indeed, (\ref{linewidth}) may not exist for the spectrum with exact Lorentzian shape. But, as will be seen later, (\ref{linewidth}) always exists for the system we are considering and it is known that (\ref{linewidth}) could be regarded as a good approximation to the width of the Lorentzian distribution in some situations~\cite{kubo_tomita,choukroun}. In this sense, we adopt (\ref{linewidth}) as the definition of the linewidth and other problems will be discussed in the last part of this section.

So far the method has been used to study the whole shape of the spectrum.
%This method was introduced in the reference~\cite{Weisse} in order to investigate the moments of the whole spectrum. 
For our purpose, however, we need to improve this method because moments of the total spectrum are not very helpful in considering the detailed shape of spectrum. 
In other words, even if 
we know moments with small $n$ of the whole spectrum, 
we could not conclude the size-dependence of the double-peak's split. 
%%%%%

However, as we saw in the previous section, the spectrum can be decomposed to contributions from transitions specified by the magnetization $M$. Thus now we extend the method and investigate properties of each contribution to obtain information for the structure of the spectrum, e.g., the double peak.
%In order to overcome this difficulty, 
We will focus on the moments of partial spectrum made by only transitions $(M\to M-1)$, instead of the whole spectrum. 
For example, if we know the mean position of the partial spectrum 
from the transitions $(0\to-1)$ (given by the solid line in Fig.~\ref{ED12_color_100}(Left)), 
we can see how the double-peak structure behaves depending on the size $N$.

From now on, we set the external field $H=0\mathrm{K}$, because the external field brings about just 
a shift of the mean position of spectra and does not affect the shape
(See Appendix A). 

The whole spectrum at infinite temperature is given by %Eqs.~(\ref{chiomega}), (\ref{DD})
\beq
\int_{-\infty}^{\infty}\langle M^xM^x(t)\rangle_{\infty}\mathrm{e}^{-\mathrm{i}\omega t}
=\frac{1}{2^N}\sum_{m,n}|\langle m|M^x|n\rangle|^{2}2\pi\delta(\omega-(E_{m}-E_{n})),
\eeq
where $\langle\cdots\rangle_{\infty}=\mathbf{Tr\,}[\cdots]/2^N$.
The partial spectrum made by transitions $(M\to M-1)$ is obtained as follows:
\begin{align}
\int_{-\infty}^{\infty}\langle P_{(M^z=M)}M^xP_{(M^z=M-1)}M^x(t)P_{(M^z=M)}\rangle_{\infty}
\mathrm{e}^{-\mathrm{i}\omega t}\mathrm{d}t
=\frac{1}{2^N}\sum_{\stackrel{m}{(M_m^z=M-1)}}\sum_{\stackrel{n}{(M_n^z=M)}}|\langle m|M^x|n\rangle|^22\pi\delta(\omega-(E_{m}-E_{n})),
\end{align}
where $P_{(M^z=M)}$ is a projection operator which projects states onto the subspace where $M^z=M$.
The spectrum from the specified transition is given by
\begin{align}
S_{xx}^{M}&\equiv\int_{-\infty}^{\infty}\langle 
P_{(M^z=M)}M^xP_{(M^z=M-1)}M^x(t)P_{(M^z=M)}\rangle_{\infty}\mathrm{e}^{-\mathrm{i}\omega t}\mathrm{d}t,\\
\end{align}
and its $n$-th order moment
\begin{align}
m_{n}^{M}&=\int_{-\infty}^{\infty}\omega^nS_{xx}^{M}(\omega)\mathrm{d}\omega\\
&=\frac{1}{2^{N}}\frac{\pi}{2}\sum_{\stackrel{\bm{\sigma}}{(M_{\bm{\sigma}}^z=M)}}\langle\bm{\sigma}|M^+\mathrm{ad}_{\mathcal{H}}^n(M^-)|\bm{\sigma}\rangle,
\label{nth_moment}
\end{align}
where $M^{\pm}\equiv M^x\pm\mathrm{i}M^y$. As for the basis set $\left\{|\bm{\sigma}\rangle\right\}$, we may use the up/down-spin representation, such as $|\uparrow\downarrow\downarrow\cdots\uparrow\downarrow\uparrow\rangle$.

Because, in the infinite-temperature limit, the Boltzmann factor does not appear, and thus Eq.~(\ref{nth_moment}) can be calculated by counting the number of state (combinatorics). 
The zeroth, first, and second order moments are explicitly written as
\begin{align}
m_{0}^{M}&=\frac{1}{2^{N}}\frac{\pi N}{2}\sum_{\stackrel{\bm{\sigma}}{(M_{\bm{\sigma}}^z=M)}}\langle\bm{\sigma}|S_1^+S_1^-|\bm{\sigma}\rangle,\\
m_{1}^{M}&=-\frac{1}{2^{N}}\pi(N-1)\Delta\sum_{\stackrel{\bm{\sigma}}{(M_{\bm{\sigma}}^z=M)}}\langle\bm{\sigma}|S_1^+S_1^-S_2^z|\bm{\sigma}\rangle,\\
m_{2}^{M}&=\frac{1}{2^{N}}\frac{\pi\Delta^2}{4}\left[(N-1)\sum_{\stackrel{\bm{\sigma}}{(M_{\bm{\sigma}}^z=M)}}\langle\bm{\sigma}|S_1^+S_1^-|\bm{\sigma}\rangle
+4(N-2)\sum_{\stackrel{\bm{\sigma}}{(M_{\bm{\sigma}}^z=M)}}\langle\bm{\sigma}|S_1^+S_1^-S_2^zS_3^z|\bm{\sigma}\rangle\right].
\end{align}
From these quantities, we obtain the intensity (the area) of the spectrum, the mean position, and the linewidth of partial spectra 
from the transitions $(M\to M-1)$ 
in the following:
\begin{align}
m_0^M&=\frac{1}{2^N}\frac{\pi N}{2}\binom{N-1}{\frac{N}{2}-M}\overset{|M|\ll N}{\sim}\sqrt{\frac{N}{2\pi}},
\label{m0}\\
\frac{m_1^M}{m_0^M}&=(1-2M)\frac{\Delta}{N},
\label{m1}\\
\sqrt{\frac{m_2^M}{m_0^M}-\left(\frac{m_1^M}{m_0^M}\right)^2}&=\frac{|\Delta|}{\sqrt{2}}\sqrt{\left(1-\frac{2}{N}\right)\left(1-\frac{(1-2M)^2}{N(N-1)}\right)}\overset{|M|\ll N}{\sim}\frac{|\Delta|}{\sqrt{2}}+\mathcal{O}\left(\frac{1}{N}\right).
\label{m2}
\end{align}
Note that these results are valid for any $\Delta$ and $J$.

Let us analyze the numerical results again from the viewpoint of Eqs.~(\ref{m0}) $\sim$  (\ref{m2}).
The heights plotted in Fig.~\ref{size_dependence} (Right) correspond to the intensity of the spectrum and they
relate to the quantity given by Eq.~(\ref{m0}).
Eq.~(\ref{m0}) indicates that the area of the partial spectrum increases with $N$ larger, which
supports the size-dependence of the peak's height.
Eq.~(\ref{m1}) is consistent with the observation of $1/N$ in the numerical results shown in Fig.~\ref{size_dependence} (Left) and (Center). 
%ikeuchi0806
We added the green lines represented as $2(1-2M)\Delta/N$ in Fig.~\ref{size_dependence}. 
There are some differences between the numerical results and the green theoretical lines because Eq.~(\ref{m1}) gives just the mean position, not the position of the maximum of the spectrum.
%Precisely speaking,
Nevertheless, the mean of the peak due to the transitions between $M=0$ and $-1$ tends to zero, which suggests that the peaks of the double peak shrink to the center.
Thus the hypothesis is strongly supported that the double peak and small protuberances also get stuck as $N\to\infty$.

As for the central peak in odd-spin systems, its mean position turns out to be always zero by substituting $M=1/2$ into Eq.~(\ref{m1}). 
%By substituting $M=1$, $3/2$ and $2$, 
%the hypothesis is strongly supported that the double peak and small protuberances also get stuck as $N\to\infty$.%, although the tops of the peaks is not necessarily found at the mean position.

Based on these facts, we can draw the following conclusion about the separation of the double peaks. The mean position of spectrum for each transition exactly goes to zero in proportional to $1/N$. This means 
%ikeuch0723
the position of 
each peak coming from the transition ($M\to M-1$), e.g., the peak shown by the solid line in Fig.~\ref{ED12_color_100}, goes to zero.
In particular, this seems to suggest that the double peak shrinks to the origin. 

\begin{figure}[H]
	\begin{center}
	\includegraphics[width=60mm]{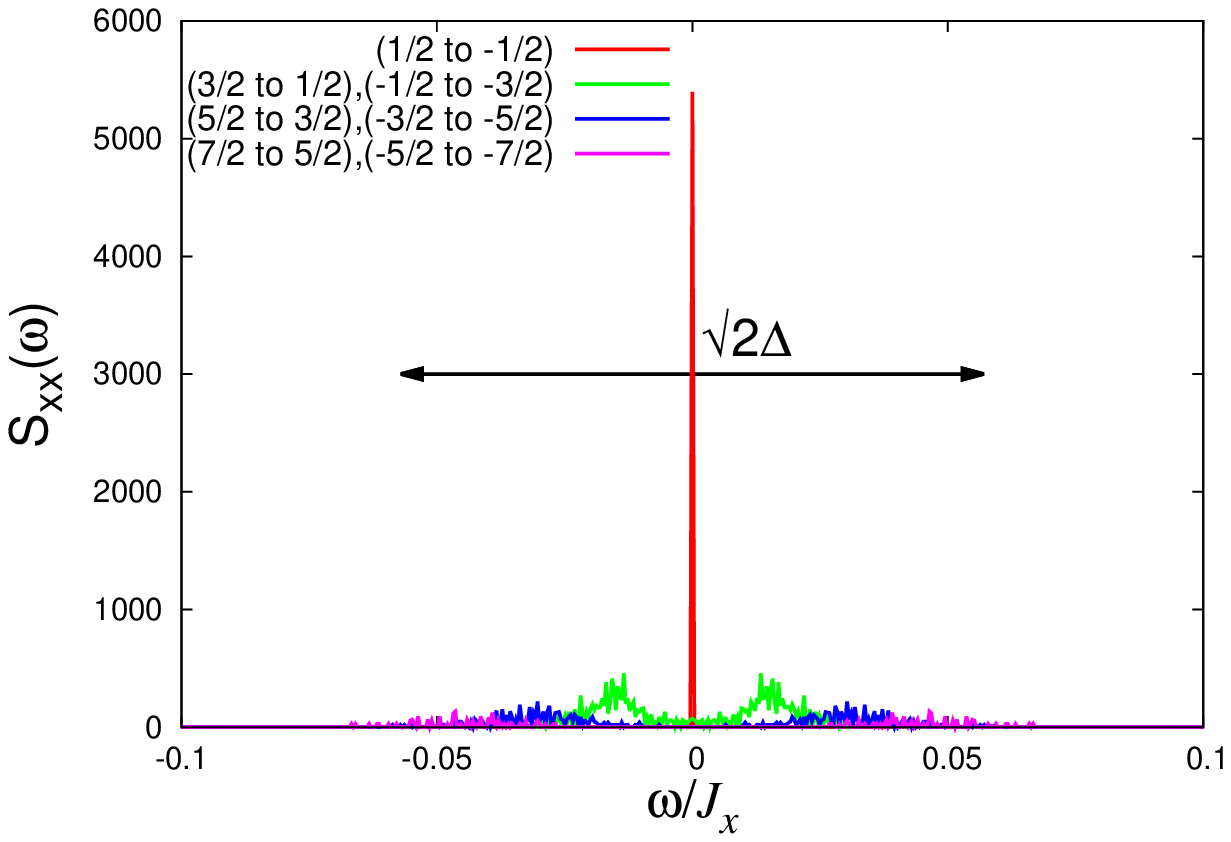}
	\includegraphics[width=60mm]{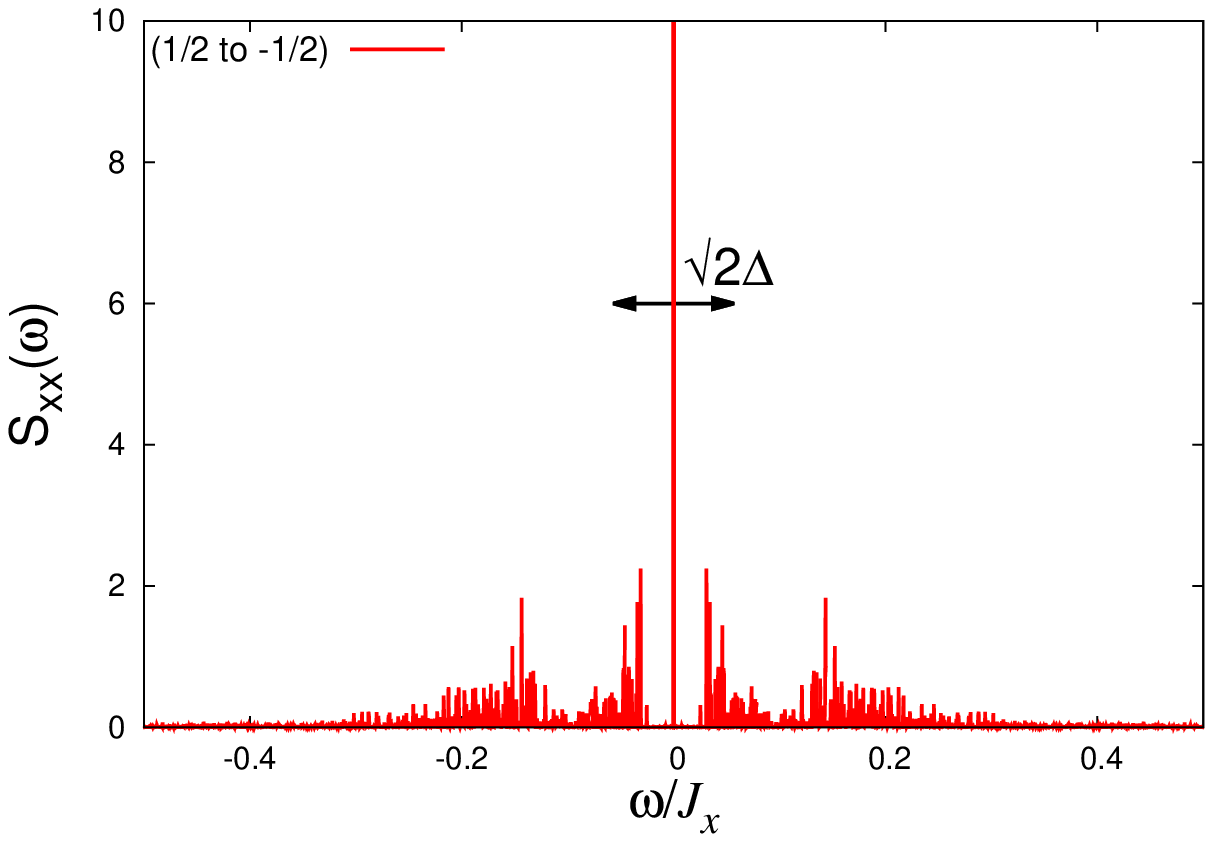}
	\end{center}
	%\vspace{10mm}
	\caption{
(Color online) (Left), (Right) Spectrum for $N=11$, $\beta^{-1}=\infty$. Note that the scales are different between (Left) and (Right). According to Eq.~(\ref{m2}), we consider $2\times\Delta/\sqrt{2}$ as the linewidth, which is denoted by the black arrow. 
}
	\label{N11Delta}
\end{figure}

On the other hand, the interpretation of Eq.~(\ref{m2}) includes a delicate problem.
For instance, let us discuss the single peak in odd-spin systems. 
In Fig.~\ref{N11Delta} (Left), $S_{xx}(\omega)$ of $N=11$ is decomposed into different values of $M$ as in the previous section. 
The linewidth of the spectrum due to the transition between  $M=\pm 1/2$ (red line) in Fig.~\ref{N11Delta}
looks very small, almost zero. 
Nevertheless, the linewidth calculated from Eq.~(\ref{m2}) is shown by the arrow. The width estimated by Eq.~(\ref{m2})
is much larger,  which seems contradictory to the data. 
In order to resolve this problem we plot the spectrum  in a different scale in which the small peaks can be seen
(Fig.~\ref{N11Delta} (Right) ). 
We find small peaks far from the central peak, which cause the linewidth broader than we expected.
In real experiments, the small peaks  in Fig.~\ref{N11Delta} (Right) may be
smeared out by noise and only the central peak can be observed. So in this sense, the linewidth calculated from Eq.~(\ref{m2}) may not be appropriate in practical situation.

With this observation, small peaks away from the origin might give unexpected contributions, and thus 
we cannot deny the possibility that the tops of the double peak do not get stuck together in the thermodynamic limit.

%\vspace{-25mm}
%
%\vspace{-20mm}

\section{Summary and Discussion}\label{sec_summary}
In the present paper, we studied the temperature- and size-dependence of the ESR spectrum for the XXZ chain. 
A drastic change from high temperature spectrum with a structure around the EPR position to the low temperature spectrum with a single peak at a shifted position from  the EPR position.
Temperature-dependence of the line shape has not been studied in detail and the drastic change has not been recognized yet in experiment. Although the structure may disappear in the thermodynamic limit as we discussed in the present paper,  the structure definitely exists up to considerable length of lattice (say $N<30$) and we hope the structure and the temperature-dependence will be observed in experiments.

Subsequently, we also investigated the size-dependence of the separation of the double peaks. The double-peak structure obtained by the AC method shows the tendency that the separation shrinks to zero as $N\to\infty$.
The size-dependence was analyzed by making use of an extended moment method. 
Since the XXZ model has a conserved quantity, 
%%0527 
i.e., the $z$-component of magnetization $M$, 
we can decompose the whole spectrum into the partial spectra specified by types of transitions.
%By focusing on the splitting of the these partial spectra due to anisotropy, we gave some explanation about the temperature-dependence of the spectra, especially the double-peak structure at high temperatures and finite shift at low temperatures.
We applied the moment method to each contribution of the spectrum from a  transition with specified magnetizations.
As the results, we found that the mean positions of all spectrum approach to the center as $\Delta/N$ and become zero as $N\to\infty$, which strongly indicates the double peak collapses at the center. The linewidth remains finite as $\Delta/\sqrt{2}$ even in the thermodynamics limit. Such observation prevents us from drowing the definite conclusion.  

Finally, we put some comments towards the future works. In this paper, since we assumed the anisotropy to be small ($\Delta/J=-0.08$) and the transitions between different multiplets can be ignorable, the decomposition of the spectrum into contribution specified by magnetization was valid. But the analytical results derived with the moment method in the last section is valid for any $J$ and $\Delta$, 
%%
%% miya0724 以下の部分は計算はしたがデータは示さないと考えていいか?
%(ikeuchi)理論的な背景がはっきり掴めておらず、生データを羅列するだけになってしまうため、
%図は示していません（数値計算はしてあります）。
%%
and besides, according to our numerical simulation, the decomposition of the spectrum specified by magnetization somehow goes well for the wide range of XY-like anisotropy ($0<\Delta/J<1$).
On the other hand, for the Ising-like anisotropy ($\Delta/J>1$), our method seems to be of limited use empirically.

The idea of the decomposition of the spectrum can be also used for other systems with magnetization conserved, e.g., systems with the Dzyaloshinskii-Moriya interaction whose direction is parallel to the static field. These applications will be studied elsewhere.

\section*{Acknowledgements}

The present work was supported by
Grants-in-Aid for Scientific Research C (25400391) from MEXT of Japan, and the Elements
Strategy Initiative Center for Magnetic Materials under the
outsourcing project of MEXT. The numerical calculations were supported by the supercomputer center of ISSP of Tokyo University.
We also acknowledge the JSPS Core-to-Core Program: Non-equilibrium dynamics of soft matter
and information.

\appendix
\section{Spectral shift with an external field $H$}
\label{shift}
In Sec. \ref{sec_separation}, we investigated the spectral shapes by calculating $S_{xx}(\omega)$ under no magnetic field, instead of
the susceptibility $\chi''(\omega)$ under a finite static field $H$. This is valid for a sufficient strong field $H\gg J(>0)$ and a sufficient high temperature $\beta\sim0$. In this appendix, we show this fact.

First, we consider the spectrum under a strong field $H$. Let the simultaneous eigenvectors of $\mathcal{H}_{0}+\mathcal{H}'$ and $\mathcal{H}_{z}$ be $\left\{|E^0_n,M_n\rangle\right\}_{n=1}^{D}$:
\begin{align}
(\mathcal{H}_{0}+\mathcal{H}')|E^0_n,M_n\rangle=E^0_n|E^0_n,M_n\rangle
\quad \mathcal{H}_{z}|E^0_n,M_n\rangle=-HM_n|E^0_n,M_n\rangle,
\quad n=1\cdots D.
\end{align}
$S_{xx}(\omega)$ is given by
\begin{align}
S_{xx}(\omega,H)=\frac{2\pi}{Z}\sum_{m,n}|\langle E^0_n,M_n|M^x|E^0_m,M_m\rangle|^2
\delta\left\{\omega-\left[(E_m-E_n)-H(M_m-M_n)\right]\right\}.
\label{S_xx}
\end{align}
Let us divide $S_{xx}(\omega,H)$ into two parts in the following:
\begin{align}
S_{xx}(\omega,H)=S^{>}_{xx}(\omega,H)+S^{<}_{xx}(\omega,H),
\label{S><}
\end{align}
where
\begin{align}
S_{xx}^{>}(\omega,H)&\equiv\frac{2\pi}{Z}\sum_{m,n}\frac{1}{4}|\langle E^0_n,M_n|M^+|E^0_m,M_m\rangle|^2
\delta\left\{\omega-\left[(E_m-E_n)-H(M_m-M_n)\right]\right\},
\label{M+}\\
S_{xx}^{<}(\omega,H)&\equiv\frac{2\pi}{Z}\sum_{m,n}\frac{1}{4}|\langle E^0_n,M_n|M^-|E^0_m,M_m\rangle|^2
\delta\left\{\omega-\left[(E_m-E_n)-H(M_m-M_n)\right]\right\},
\label{M-}
\end{align}
%
%%0527
and we find
\begin{align}
S_{xx}^{>}(-\omega,H)=S_{xx}^{<}(\omega,H).
\label{M+-}
\end{align}
Noting that we are interested in the absorption, not the emission, we may focus on the region $\omega>0$. The peaks in this region need to satisfy the relation $E_m-E_n>H(M_m-M_n)$, then it follows that $M_m=M_n-1$, because $H\gg J>0$.
Therefore, only $S^{>}_{xx}(\omega,H)$ contributes to the spectrum $S_{xx}(\omega,H)$ in the region $\omega>0$:
\begin{align}
S_{xx}(\omega,H)=S^{>}_{xx}(\omega,H),\quad\omega>0,
\label{omega+}\\
S_{xx}(\omega,H)=S^{<}_{xx}(\omega,H),\quad\omega<0.
\end{align}

According to Eqs.~(\ref{omega+}) and (\ref{M+}), we have $S_{xx}(\omega,H)=S^{>}_{xx}(\omega-H,0)$. On the other hand, it is shown that
\begin{align}
S^{>}_{xx}(\omega-H,0)=S^{>}_{xx}(H-\omega,0)=S^{<}_{xx}(\omega-H,0),
\end{align}
because of the symmetry of the shape of $S^{>}_{xx}$ as seen in Sec.~\ref{sec_decomp}, and Eq.~(\ref{M+-}). Then, by using Eq.~(\ref{S><}), we have $S^{>}_{xx}(\omega-H,0)=\frac{1}{2}S_{xx}(\omega-H,0)$. Therefore, it follows that $S_{xx}(\omega,H)=\frac{1}{2}S_{xx}(\omega-H,0)$, which is what we wanted to show.

The difference between $S_{xx}(\omega)$ and $\chi''(\omega)$ is just the presence of the factor $(1-\mathrm{e}^{-\beta\omega})/2\sim\beta\omega/2$. $\chi''(\omega)$ vanishes at infinite temperature because of this factor, but we are interested in the spectral shape, not in the exact value of the peak. So we can ignore this factor and consider only $S_{xx}(\omega)$. Strictly speaking, the $\omega$-dependence of the factor $\beta\omega/2$ could deform the spectral shape, but this effect is also ignorable in the case where $H$ is very large.

\end{document}